\documentclass{emulateapj}

\usepackage{natbib}
\usepackage{hyperref}
\usepackage{color}
\usepackage{xspace}
 \usepackage{amsmath}
\shorttitle{Evolution of the internal structure of massive galaxies}
\shortauthors{Sonnenfeld et~al.}

\def\ucsb{1}
\def\iap{2}
\def\kipac{3}
\def\asiaa{4}
\def\oxford{5}
\def\cambridge{6}
\def\bologna{7}

\def\rein{R_{\mathrm{Ein}}}
\def\reff{R_{\mathrm{eff}}}

\def\sigmae2{\sigma_{\rm e2}}
\def\meangamma{\langle \gamma' \rangle}
\def\pr{{\rm Pr}}

\newcommand{\kms}{\, {\rm km\, s}^{-1}}

\def\Sref#1{Section~\ref{#1}\xspace}
\def\Fref#1{Figure~\ref{#1}\xspace}
\def\Tref#1{Table~\ref{#1}\xspace}
\def\Eref#1{Equation~\ref{#1}\xspace}

\begin{document}

\title{The SL2S Galaxy-scale Lens Sample.  IV. 
The dependence of the total mass density profile of early-type galaxies on redshift, stellar mass, and size}

\author{Alessandro~Sonnenfeld\altaffilmark{\ucsb}$^{*}$}
\author{Tommaso~Treu\altaffilmark{\ucsb}$^{\dag}$}
\author{Rapha\"el~Gavazzi\altaffilmark{\iap}}
\author{Sherry~H.~Suyu\altaffilmark{\ucsb,\kipac,\asiaa}}
\author{Philip~J.~Marshall\altaffilmark{\oxford,\kipac}}
\author{Matthew~W.~Auger\altaffilmark{\cambridge}}
\author{Carlo~Nipoti\altaffilmark{\bologna}}
%\author{Vardha~N.~Bennert\altaffilmark{\datapeople}}
%\author{Marusa~Bradac\altaffilmark{\datapeople}}
%\author{Florence~Brault\altaffilmark{\iap}}
%\author{the SL2S collaboration}

% Auger, Sonnenfeld, Treu, Suyu:
\altaffiltext{\ucsb}{Physics Department, University of California, Santa Barbara, CA 93106, USA} 
% Gavazzi, Brault:
\altaffiltext{\iap}{Institut d'Astrophysique de Paris, UMR7095 CNRS - Universit\'e
 Pierre et Marie Curie, 98bis bd Arago, 75014 Paris, France}
% Suyu
\altaffiltext{\asiaa}{Institute of Astronomy and Astrophysics, Academia Sinica, P.O.~Box 23-141, Taipei 10617, Taiwan}
\altaffiltext{\kipac}{Kavli Institute for Particle Astrophysics and Cosmology, Stanford University, 452 Lomita Mall, Stanford, CA 94305, USA}
% Marshall:
\altaffiltext{\oxford}{Department of Physics, University of Oxford, Keble Road, Oxford, OX1 3RH, UK}
%Vardha and Marusa
%\altaffiltext{\datapeople}{They took data for us. Keep them?}
\altaffiltext{\cambridge}{Institute of Astronomy, University of Cambridge, Madingley Rd, Cambridge, CB3 0HA, UK}
\altaffiltext{\bologna}{Astronomy Department, University of Bologna, via Ranzani 1, I-40127 Bologna, Italy}

\altaffiltext{*}{{\tt sonnen@physics.ucsb.edu}}
\altaffiltext{$\dag$}{{Packard Research Fellow}}

%-------------------------------------------------------------------------------

\begin{abstract}

We present optical and near infrared spectroscopy obtained at
Keck, VLT, and Gemini for a sample of 36 secure strong gravitational
lens systems and 17 candidates identified as part of the SL2S survey.
The deflectors are massive early-type galaxies in the redshift range
$z_d=0.2-0.8$, while the lensed sources are at $z_s=1-3.5$. We combine
this data with photometric and lensing measurements presented in the
companion paper III and with lenses from the SLACS and LSD surveys to
investigate the cosmic evolution of the internal structure of massive
early-type galaxies over half the age of the universe.  We study the
dependence of the slope of the total mass density profile $\gamma'$
($\rho(r)\propto r^{-\gamma'}$) on stellar mass, size, and
redshift. We find that two parameters are sufficent to determine
$\gamma'$ with less than 6\% residual scatter. At fixed redshift,
$\gamma'$ depends solely on the surface stellar mass density $\partial
\gamma'/ \partial \Sigma_*=0.38\pm 0.07$, i.e. galaxies with denser stars 
also have steeper slopes. At fixed $M_*$ and $\reff$, $\gamma'$
depends on redshift, in the sense that galaxies at a lower redshift
have steeper slopes ($\partial \gamma' / \partial z = -0.31\pm
0.10$). However, the mean redshift evolution of $\gamma'$ for an individual
galaxy is consistent with zero $\mathrm{d}\gamma'/\mathrm{d}z=-0.10\pm0.12$. This result
is obtained by combining our measured dependencies of $\gamma'$ on
$z,M_*$,$\reff$ with the evolution of the $\reff$-$M_*$ taken from the
literature, 
%Qualitatively, our observations are consistent with a
%scenario where individual galaxies grow in stellar mass and decrease
%in density since $z=1$ -- perhaps driven by the accretion of lower
%density material via minor dry mergers -- while substantially
%preserving their total mass density profile. The apparent evolution at
%fixed M$_*$ and $\reff$ is due to the fact that the denser galaxies have also
%steeper mass density profiles, possibly as a result of higher star
%formation efficiencies. 
and is broadly consistent with current models
of the formation and evolution of massive early-type
galaxies. Detailed quantitative comparisons of our results with theory
will provide qualitatively new information on the detailed physical
processes at work.
\end{abstract}


%-------------------------------------------------------------------------------

\section{Introduction}\label{sect:intro}

The formation and evolution of early-type galaxies (ETGs) is still an
open question.  Though frequently labeled as ``red and dead'' and
traditionally thought to form in a ``monolithic collapse'' followed by
``passive'' pure luminosity evolution, in the past decade a far more
complicated history has emerged \citep[e.g.,][and references
therein]{Ren06}. ETGs are thought to harbor supermassive black holes
at their centers \citep{F+M00,Geb++00} which regulate the conversion
of gas into stars \citep{DeL++06}. Traces of recent star formation are
ubiquitously found when sensitive diagnostics are applied
\citep{Tre++02,Kav10}. Episodes of tidal disturbances and interactions
with other systems occur with remarkable frequency even at recent times
\citep[e.g.][]{m+c83,pvd05,Tal++09,Atk++13}. Their structural
properties evolve in the sense that their sizes appear to grow with
time at fixed stellar mass \citep{vDo++08,Dam++11,New++12,Hue++13,Car++13}.
The mode of star formation seems to be different from that found in spiral
galaxies, resulting in a different stellar initial mass function
\citep{Tre++10,v+C10,Aug++10b,Bre++12,Cap++13}. Finally, from a
demographic point of view, their number density has been found to have
evolved significantly since $z\sim2$ \citep[e.g.,][]{Ilb++13}.

Reproducing these observations is an enormous challenge for
theoretical models. Major and minor mergers are thought to be the
main processes driving their structural and morphological evolution,
but it is not clear if they can account for the observed evolution
while reproducing all the observables
\citep{Nip++09,Hop++10d,Ose++12,Rem++13}. 
%Major and minor mergers
%affect the luminous and dark matter components in different ways
%\citep{Nip++12}, therefore disentangling the two components is a
%powerful diagnostic tool.  Although many different studies have
%described the evolution of the stellar component of ETGs
%\citep{Fon++04,Cim++06,Poz++10,Pen++10}, very little is known about the
%evolution of their dark matter halos from traditional tracers of mass
%like stellar kinematics \citep{vdM+vD07,Cap++09}.  

Gravitational lensing, by itself and in combination with other probes,
can be used to great effect to measure the 
mass profiles of early-type galaxies, both in the 
nearby universe and at cosmological distances
%separate the luminous and dark matter
%components of early-type galaxies in nearby universe and at
%cosmological distances
\citep{T+K02a,T+K02b,RKK03,T+K04,R+K05,Koo++06,J+K07,Gav++07,Aug++10,Lag++10}.
Until recently, however, this approach was severely limited by the
small size of the known samples of strong gravitational lenses. This
has motivated a number of dedicated searches which have,  in the past
decade, increased the sample of known strong gravitational lens systems
by more than an order of magnitude
\citep[e.g.,][]{Bro++03,Bol++08,Fau++08,Tre++11}.

In spite of all this progress the number of known lenses at $z\sim0.5$
and above is still a severe limitation. Increasing this sample and
using it as a tool to understand the formation and evolution of massive
galaxies is the main goal of our SL2S galaxy-scale lens search
\citep{Gav++12} and other independent searches based on a
variety of methods
\citep{Bro++12,Mar++09,Neg++10,Paw++12,Ina++12,Gon++12,War++13,Vie++13}.

In our pilot SL2S paper \citep{Ruf++11} we measured the evolution of
the density slope of massive early-type galaxies by combining lensing
and dynamics measurements of a sample of just 11 SL2S lenses 
with similar
measurements taken from the literature \citep{T+K04,Koo++09,Aug++10b},
finding
tentative evidence that the density profile of massive ETGs steepens
with cosmic time on average.  This trend was later confirmed
qualitatively by an independent study of \citet{Bol++12} and agrees
with the theoretical work by \citet{Dub++13}. However, the picture is
not clear:  the observed trend is tentative at best, while different
theoretical studies find contrasting evolutionary trends
\citep{JNO++12,Rem++13}. More data and better models are needed to
make progress.

In order to clarify the observational picture, we have collected a
much larger sample of objects, more than tripling the sample of secure
lenses with all the necessary measurements, with respect to our pilot
study. Photometric and strong lensing measurements for this expanded
sample are presented in a companion paper \citep[hereafter Paper III]{PaperIII}.
In this paper we present spectroscopic data for the
same objects. Deflector and source redshifts are used to convert the
geometry of the lens system into measurements of a physical mass
within a physical aperture. Stellar velocity dispersions are used as
an independent constraint on the gravitational potential of the lens,
allowing for more diagnostic power on the structure of our targets. 

The combination of the photometric, lensing, and spectroscopic data is
used in this paper to study the cosmic evolution of the slope of the
average
mass density profile of massive early-type galaxies. This is achieved
by fitting power law density profiles ($\rho(r) \propto r^{-\gamma'}$;
$\gamma'\approx2$ in the local universe) to the measured Einstein
radii and velocity dispersions of our lenses.
Such a measurement of $\gamma'$ is a good proxy for the 
mean density slope within the effective radius.
The goal of this paper is to measure trends of $\gamma'$ with 
redshift, in continuity with our previous work \citep{Ruf++11}, 
as well as with other structural properties of massive ETGs, 
such as stellar mass and size.
Such measurements will help us understand the structural evolution 
of ETGs from $z=0.8$ to present times.

% A major difference with respect to Paper II is that the current size of
% our sample allows us to measure a trend with redshift of the density
% slope $\gamma'$ by using data from our survey alone.  This is very
% important because it allows us to probe the evolution of $\gamma'$ on a
% set of lenses chosen with the same selection function, thus avoiding
% possible biases that might be introduced when combining data from
% different surveys.

This paper is organized as follows. We briefly summarize the relevant
features of the SL2S galaxy scale lens sample in \Sref{sect:sample}, and
show in detail the spectroscopic data set and the measurements of
redshifts and velocity dispersions of our lenses in \Sref{sect:spec}. In
\Sref{sect:class} we discuss the properties of SL2S lenses in relation
with lenses from independent surveys.  In \Sref{sect:gammap} we briefly
explain how lensing and kinematics measurements are combined to infer
the density slope $\gamma'$ and discuss the physical meaning of such
measurements.  In \Sref{sect:euler} we combine individual $\gamma'$
measurements to infer trends of this parameter across the population of
ETGs.  After a discussion of our results in \Sref{sect:discuss} we
conclude in \Sref{sect:concl}.

Throughout this paper magnitudes are given in the AB system.  We
assume a concordance cosmology with matter and dark energy density
$\Omega_m=0.3$, $\Omega_{\Lambda}=0.7$, and Hubble constant $H_0$=70
km s$^{-1} $Mpc$^{-1}$.

%-------------------------------------------------------------------------------

\section{The sample}\label{sect:sample}

The gravitational lenses studied in this paper were discovered as part
of the Strong Lensing Legacy Survey \citep[SL2S]{Cab++07} with a
procedure described in detail in \citet{Gav++12}. Lens candidates are
identified in imaging data from the CFHT Legacy Survey and then
followed up with \textit{Hubble Space Telescope} (\textit{HST}) high
resolution imaging and/or
spectroscopy. In Paper III we ranked the candidates, assigning them a
grade indicating their likelihood of being strong lenses, with the
following scheme: grade A for definite lenses, grade B for probable
lenses, grade C for possible lenses and grade X for non-lenses. A
summary with the number of systems in each category is given in
\Tref{table:census}. In this paper we analyze all lenses with
spectroscopic data that have not been ruled out as grade X systems.
\begin{deluxetable}{lccccc}
 \tablecaption{ \label{table:census} Census of SL2S lenses.}
 \tablehead{
Grade & A & B & C & X & Total }
 \startdata
 With high-res imaging & 30 & 3 & 13 & 21 & 67\\ 
With spectroscopy & 36 & 15 & 2 & 5 & 58 \\ 
High-res imaging and spectroscopy & 27 & 3 & 0 & 0 & 30 \\ 
Total with follow-up & 39 & 15 & 15 & 26 & 95 \\ 

 \enddata
 \tablecomments{Number of SL2S candidates for which we obtained
   follow-up observations in each quality bin. Grade A: definite
   lenses, B: probable lenses, C: possible lenses, X:
   non-lenses. We differentiate between lenses with spectroscopic
   follow-up, high-resolution imaging follow-up or any of the two.}
\end{deluxetable}

%-------------------------------------------------------------------------------

\section{Spectroscopic observations}\label{sect:spec}

The SL2S spectroscopic campaign was started in 2006. 
The goal of our spectroscopic observations is to measure the lens and source redshifts and lens velocity dispersion for all our systems.
Different telescopes (Keck, VLT and Gemini), instruments (LRIS, DEIMOS, X-Shooter, GNIRS) and setups have been used to achieve this goal, reflecting technical advances during the years and the optimization of our strategy.
In what follows we describe the procedure used to measure the three key spectroscopic observables. 
A summary of all measurements is given in \Tref{table:allspec}.

% - - - - - - - - - - - - - - - - - - - - - - - - - - - - - - - - - - - - - - - 

\subsection{Deflector redshifts and velocity dispersions}\label{ssect:zd}

The typical brightness of our lenses is around $i\sim20$. With an 8m class telescope, their redshift can be measured from their optical absorption lines with $\sim10$ minutes of exposure time, while a measurement of their velocity dispersion typically takes from 30 to 120 minutes.
Optical spectroscopy data come from three different instruments. %: the LRIS and DEIMOS spectrographs at Keck and XSHOOTER at VLT.

For most of the systems we have data obtained with the LRIS spectrograph at Keck \citep{Oke++95}.
The wavelength coverage of LRIS is typically in the range $3500-8000$ \AA~for data taken before 2009 and extends up to $10000$ \AA~for later data, after the installation of the new detector with much reduced fringing patterns \citep{Roc++10}. The spectral resolution is about 140 km s$^{-1}$ FWHM on the red side of the spectrograph.
%In the early years of our campaign, a variety of setups was used \ref{Ruf++11}.
%Later on, the typical configuration consisted of a 1.0'' slit, the 600/7500 red grating, 400/6000 blue grism and 680 dichroic. 
%The spectral resolution of this setup is about 60 km s$^{-1}$ in the red, more than adequate to measure velocity dispersions of massive galaxies. 
%Taking advantage of the new LRIS detector installed in 2009, this configuration allows us to cover a wavelength range from $3500\AA$ up to $10000\AA$.
%One drawback is that for many of our objects the most prominent absorption feature, the Ca K,H doublet, falls on the blue detector, which does not have a good enough spectral resolution.
%This however does not prevent us from measuring velocity dispersions of those galaxies, as other lines at redder wavelength are used.
Data reduction for LRIS spectra was performed with a pipeline written by M.W. Auger.

For a set of 13 systems we have VLT observations with the instrument X-Shooter\footnote{ESO/VLT programs 086.B-0407(A) and 089.B-0057(A), PI Gavazzi}.
X-Shooter has both a higher resolution ($\sim50\kms$) and a longer wavelength coverage (from 3500 \AA~up to 25000 \AA) than LRIS.
X-Shooter spectra were reduced with the default ESO pipeline\footnote{\url{http://www.eso.org/sci/facilities/paranal/instruments/xshooter/}}. The observations were done by nodding along a long slit of width $0\farcs9$ for the UVB and VIS arms and $1\farcs0$ for the NIR arm.

Finally, six systems presented here have data obtained with DEIMOS at Keck \citep{Fab++03}.
The grating used in all DEIMOS observations is the 600ZD, with a wavelength range between 4500 \AA~and 9500 \AA~and a spectral resolution of about $160\kms$.
DEIMOS data were reduced with the DEEP2 pipeline \citep{Coo++12,New++12}.

Both redshifts and velocity dispersions are measured by fitting stellar templates, broadened with a velocity kernel, to the observed spectra.
This is done in practice with a Monte Carlo Markov Chain adaptation of the velocity dispersion fitting code by \citet{vandermarel1994}, written by M. W. Auger and described by \citet{Suy++10}.
We used 7 different templates of G and F stars, which should provide an adequate description of the stars in red passive galaxies, taken from the Indo US stellar library. 
The code also fits for an additive polynomial continuum, to accomodate for template mismatch effects or imperfections in the instrumental response correction.
In most cases, a polynomial of order five is used.

The rest-frame wavelength range typically used in our fits is $3850\AA-5250\AA$, which brackets important absorption lines such as Ca K,H at $3934,3967\AA$, the G-band absorption complex around $4300\AA$ and Mgb at $5175\AA$.
Depending on the redshift of the target and the instrument used, this is not always allowed as part of the wavelength region can fall outside the spectral coverage allowed by the detector, or because of Telluric absorption.
In those cases the fitted rest-frame region is extended.
%For example, for the spectra taken with LRIS in the years 2010-2011 a low-resolution blue grism was used, which does not allow for reliable velocity dispersion measurements at observed wavelengths below $6700\AA$. Velocity dispersions are then measured with data from the red detector in the rest-frame wavelength region up to $6000\AA$, including the NaD $5892\AA$ line.

Systematic uncertainties in the velocity dispersion measurements are estimated by varying the fitted wavelength region and order of the polynomial continuum. These are typically on the order of $20\kms$ and are then summed in quadrature to the statistical uncertainty.

All the optical spectra of our systems are shown in \Fref{fig:spec}.

% - - - - - - - - - - - - - - - - - - - - - - - - - - - - - - - - - - - - - - - 

\subsection{Source spectroscopy}\label{ssect:zs}

Measuring the redshift of a lensed background source is important not only for determining the geometry of the gravitational lens system, but also to confirm that the arc is actually in the background relative to the lens.
The arcs of the lensed sources are relatively faint in broad band photometry ($g\sim24$), implying that their continuum radiation cannot be detected in most cases.
However the sources are selected to be blue \citep{Gav++12} and are often associated with emission lines from star formation and/or nuclear activity.
The typical redshifts of our arcs are in the range $1<z<3$. This means that optical spectroscopy can effectively detect emission from the [OII] doublet at $3727-3729\AA$, for the lowest redshift sources, or Ly-$\alpha$ for objects at $z>2.5$ or so.
This is the case for roughly half of the systems observed. The remaining half does not show detectable emission line in the observed optical part of the spectrum, either because the most important lines fall in the near-infrared, or because emission is too weak.
Emission lines from the arcs can be easily distinguished by features in the lens because they are spatially offset from the lens light.
%Spatial information can be a decisive factor in the confirmation of a gravitational lens system: a line emission that is multiply imaged on both sides of the foreground object is a very convincing evidence of the lens nature of the system.

X-Shooter observations proved to be remarkably efficient in measuring source redshifts. This is in virtue of its wavelength range that extends through the near infrared up to $25000\AA$ and its medium resolution that limits the degrading effect of emission lines from the atmosphere.
Of 13 systems observed with X-Shooter, 12 of them yielded a redshift of the background source, all of which with at least two identified lines.

% SHS: I have moved this paragraph upward (it used to follow the next long
% paragraph)
In addition, for four systems we have near infrared spectroscopic observations with the instrument GNIRS on Gemini North (PI Marshall, GN-2012B-Q-78, PI Sonnenfeld, GN-2013A-Q-91), used in cross-dispersed mode, covering the wavelength range $10000\AA-25000\AA$ at once.
Of the four systems observed, two of them show two emission lines from the background source.

In most cases when only optical spectroscopy is available, only one emission line is detected over the whole spectrum.
The [OII] doublet can be easily identified even with relatively low resolution spectrographs. %Different is the case for Ly-$\alpha$ emission.
% SHS revised:
The identification of the Ly-$\alpha$ line is less trivial. 
Ly-$\alpha$ is typically the brightest emission line in the rest frame wavelength range $1000-3000\AA$ when present, but other emission lines like CIV 1546\AA, OIII 1666\AA or CIII 1908\AA~can sometimes be seen.
When we detect an emission line close to the blue end of the spectrum it could in principle be any of those lines. However a detection of one of the above lines and a non-detection of the other ones is quite unlikely, unless CIII 1908\AA~falls right at the blue edge of the observed spectrum. In that case though we should expect to observe the OII doublet at redder wavelengths. This case is never encountered, therefore in all cases when we detect an unresolved emission line bluer than $6000\AA$, and no other lines, we can safely assume it is Ly-$\alpha$. 
The system SL2SJ022357-065142 is a particular case: we detected an emission line spatially associated with the background source at $9065\AA$, with a $5-\sigma$ significance. Given the low S/N the line is both compatible with being the OII doublet or an individual line. Possible other lines are OIII 5007\AA and H-$\beta$, which cannot be ruled out. Therefore we do not claim redshift measurements for that source: deeper data is needed to establish whether the line is the OII doublet or not.

% SHS: I have rearranged/split the paragraph below:
%Finally, six background sources are bright enough to be visible with continuum radiation and several absorption/emission features can be identified in their spectra.
%2d spectra around all the detected emission lines for all the systems are shown in \Fref{fig:emline} and the absorption line spectra of the few sources that show them are plotted in \Fref{fig:absorption}.
%Note how for some systems the line emission is multiply imaged on both sides of the foreground object. This provides a decisive clue on the lens nature of those systems, important when ranking our targets by their likelihood of being lenses (Paper III).

\begin{figure*}
\centering\includegraphics[width=0.9\textwidth]{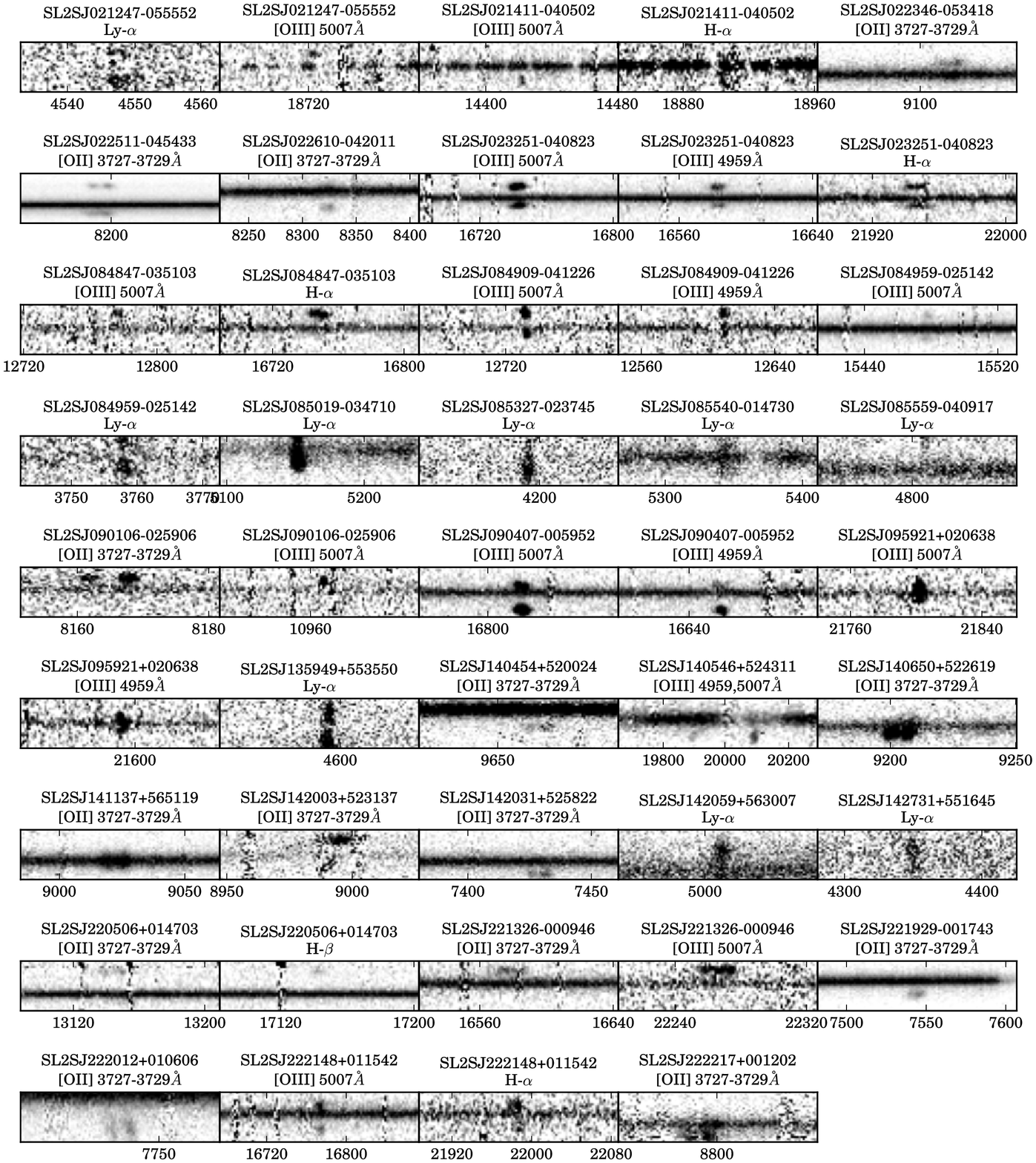}
\caption{\label{fig:emline} 2d spectra of SL2S lenses around the identified emission lines from the lensed arcs. Observer frame wavelength in \AA is labeled on the horizontal axis.}
\end{figure*}

The 2d spectra around all the detected emission lines for all the systems are shown in \Fref{fig:emline}.  Note that for some systems the line emission is multiply imaged on both sides of the foreground object. This provides a decisive clue on the lens nature of those systems, important when ranking our targets by their likelihood of being lenses (Paper III).  

\begin{figure*}
\centering\includegraphics[width=0.9\textwidth]{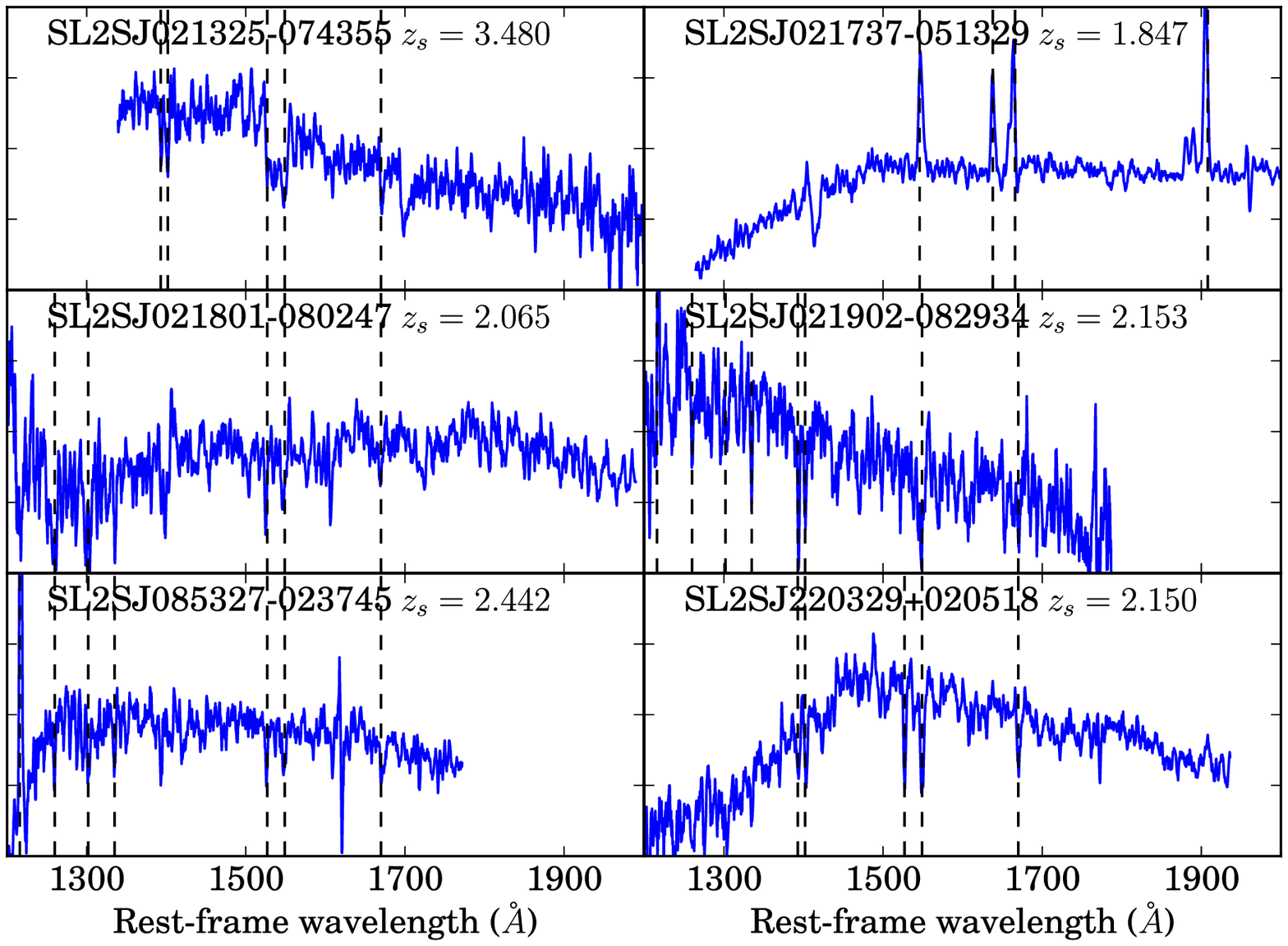}
\caption{\label{fig:absorption} Spectra of lensed sources that are bright enough to be detected in the continuum. The vertical dashed lines highlight absorption/emission line features: in order of increasing wavelength Ly-$\alpha$ (1216\AA), SiII (1260\AA), SiII (1302\AA, 1304\AA), CII (1335\AA), SiIV (1393\AA, 1402\AA), SiII (1527\AA), CIV (1549\AA), AlII (1670\AA).}
\end{figure*}

Finally, six background sources are bright enough to be visible with continuum radiation and several absorption/emission features can be identified in their spectra.  The absorption line spectra of these sources are plotted in \Fref{fig:absorption}.

Despite our efforts in acquiring spectroscopic data for our lenses, seven of the 36 grade A lenses with spectroscopic follow-up have no measured source redshifts.
In Paper II \citet{Ruf++11} made use of photometric data together with lensing cross-section arguments to estimate source redshifts, with a technique called {\em photogeometric redshift}.
Here the fraction of lenses with no source redshift is small compared to the sample size, therefore it is not essential to include them in the analysis through the use of this method.
%A follow-up study on the applicability of photogeometric redshift estimates is shown in the Appendix.
% - - - - - - - - - - - - - - - - - - - - - - - - - - - - - - - - - - - - - - - 

% %%%%%%%%%%%%%%%%%%%%%%%%%%%%%%%%%%%%%
\renewcommand{\arraystretch}{1.10} 
\begin{deluxetable*}{lccccccccccc}
\tablewidth{0pt}
\tablecaption{\label{table:allspec} Spectroscopic observations.}
\tabletypesize{\scriptsize}
\tablehead{
\colhead{Name} & \colhead{obs. date} & \colhead{Instrument} & 
\colhead{slit} & \colhead{width} & \colhead{seeing} & \colhead{exp. time} & \colhead{$z_d$} & \colhead{$z_s$} 
& \colhead{$\sigma$} & \colhead{S/N} & \colhead{res.} \\
& & & ($''$) & ($''$) & ($''$) & (s) & & & (km/s) & (\AA$^{-1}$) & 
(km/s)
}
\startdata
SL2SJ020833-071414 & 11-29-2011 & LRIS & 1.0 & 1.62 & 1.0 & 900 & 0.428 & $\cdots$ & $295 \pm 27$ & 17 & $150$ \\ 
SL2SJ021206-075528 & 01-28-2011 & LRIS & 0.7 & 1.62 & 0.6 & 2700 & 0.460 & $\cdots$ & $257 \pm 25$ & 28 & $120$ \\ 
SL2SJ021247-055552 & 10-08-2010 & XSHOOTER & 0.9 & 1.60 & 0.7 & 2800 & 0.750 & 2.74 & $273 \pm 22$ & 22 & $47$ \\ 
  & 12-09-2012 & DEIMOS & 1.0 & 1.90 & 1.2 & 3600 &   &   & $253 \pm 28$ & 11 & $170$ \\ 
SL2SJ021325-074355 & 09-14-2007 & LRIS & 1.0 & 1.68 & 0.6 & 1800 & 0.717 & 3.48 & $293 \pm 34$ & 5 & $220$ \\ 
SL2SJ021411-040502 & 12-09-2012 & DEIMOS & 1.0 & 1.88 & 0.8 & 3600 & 0.609 & 1.88 & $287 \pm 47$ & 10 & $170$ \\ 
  & 01-28-2011 & LRIS & 0.7 & 1.62 & 0.6 & 2700 &   &   & $264 \pm 26$ & 13 & $120$ \\ 
  & 10-08-2010 & XSHOOTER & 0.9 & 1.60 & 0.7 & 2800 &   &   & $209 \pm 20$ & 27 & $49$ \\ 
SL2SJ021737-051329 & 12-23-2006 & LRIS & 1.5 & 1.68 & 0.6 & 2400 & 0.646 & 1.85 & $239 \pm 27$ & 11 & $160$ \\ 
  & 09-14-2007 & LRIS & 1.0 & 1.68 & 0.6 & 3600 &   &   & $292 \pm 33$ & 12 & $120$ \\ 
SL2SJ021801-080247 & 01-28-2011 & LRIS & 0.7 & 1.62 & 0.6 & 1800 & $\cdots$ & 2.06 & $\cdots$ & 6 & $120$ \\ 
  & 12-09-2012 & DEIMOS & 1.0 & 0.81 & 1.0 & 1200 &   &   & $\cdots$ & 5 & $170$ \\ 
SL2SJ021902-082934 & 09-13-2007 & LRIS & 1.0 & 1.68 & 0.7 & 2700 & 0.389 & 2.15 & $289 \pm 23$ & 21 & $210$ \\ 
SL2SJ022046-094927 & 12-09-2012 & DEIMOS & 1.0 & 1.90 & 0.8 & 1800 & 0.572 & $\cdots$ & $254 \pm 29$ & 10 & $170$ \\ 
SL2SJ022056-063934 & 09-13-2007 & LRIS & 1.0 & 1.68 & 0.8 & 1800 & 0.330 & $\cdots$ & $231 \pm 25$ & 23 & $220$ \\ 
SL2SJ022346-053418 & 11-30-2011 & LRIS & 1.0 & 1.62 & 0.6 & 900 & 0.499 & 1.44 & $288 \pm 28$ & 20 & $140$ \\ 
SL2SJ022357-065142 & 08-06-2010 & LRIS & 1.0 & 1.64 & 1.0 & 900 & 0.473 & $\cdots$ & $312 \pm 27$ & 23 & $160$ \\ 
  & 11-01-2010 & LRIS & 1.0 & 1.64 & 0.9 & 900 &   &   & $289 \pm 28$ & 25 & $150$ \\ 
SL2SJ022511-045433 & 09-09-2009 & LRIS & 1.0 & 0.81 & 0.7 & 1800 & 0.238 & 1.20 & $234 \pm 21$ & 54 & $500$ \\ 
SL2SJ022610-042011 & 09-14-2007 & LRIS & 1.0 & 1.62 & 0.6 & 1800 & 0.494 & 1.23 & $263 \pm 24$ & 15 & $230$ \\ 
SL2SJ022648-040610 & 12-23-2006 & LRIS & 1.5 & 1.68 & 0.6 & 2700 & 0.766 & $\cdots$ & $333 \pm 24$ & 9 & $160$ \\ 
  & 10-08-2010 & XSHOOTER & 0.9 & 1.60 & 0.6 & 2800 &   &   & $324 \pm 21$ & 43 & $47$ \\ 
SL2SJ022648-090421 & 09-14-2007 & LRIS & 1.0 & 1.68 & 0.6 & 1800 & 0.456 & $\cdots$ & $302 \pm 24$ & 23 & $220$ \\ 
SL2SJ023251-040823 & 09-13-2007 & LRIS & 1.0 & 1.68 & 0.7 & 2700 & 0.352 & 2.34 & $281 \pm 26$ & 19 & $220$ \\ 
  & 10-06-2010 & XSHOOTER & 1.0 & 1.60 & 0.7 & 2800 &   &   & $247 \pm 32$ & 37 & $49$ \\ 
SL2SJ084847-035103 & 01-03-2011 & XSHOOTER & 0.9 & 1.60 & 1.0 & 2800 & 0.682 & 1.55 & $197 \pm 21$ & 19 & $49$ \\ 
SL2SJ084909-041226 & 01-02-2011 & XSHOOTER & 0.9 & 1.60 & 0.9 & 2800 & 0.722 & 1.54 & $320 \pm 24$ & 14 & $49$ \\ 
  & 12-09-2012 & DEIMOS & 1.0 & 1.88 & 0.8 & 6000 &   &   & $275 \pm 26$ & 26 & $160$ \\ 
SL2SJ084934-043352 & 01-28-2011 & LRIS & 0.7 & 1.62 & 0.6 & 1800 & 0.373 & $\cdots$ & $245 \pm 24$ & 23 & $120$ \\ 
SL2SJ084959-025142 & 01-01-2011 & XSHOOTER & 0.9 & 1.60 & 0.8 & 2800 & 0.274 & 2.09 & $276 \pm 35$ & 67 & $47$ \\ 
SL2SJ085019-034710 & 01-28-2011 & LRIS & 0.7 & 1.62 & 0.6 & 2700 & 0.337 & 3.25 & $290 \pm 24$ & 26 & $120$ \\ 
SL2SJ085327-023745 & 11-30-2011 & LRIS & 1.0 & 1.62 & 0.9 & 4800 & 0.774 & 2.44 & $\cdots$ & $\cdots$ & $150$ \\ 
SL2SJ085540-014730 & 01-28-2011 & LRIS & 0.7 & 1.62 & 0.6 & 3600 & 0.365 & 3.39 & $222 \pm 25$ & 24 & $120$ \\ 
  & 12-09-2012 & DEIMOS & 1.0 & 1.88 & 0.8 & 2400 &   &   & $209 \pm 31$ & 14 & $160$ \\ 
SL2SJ085559-040917 & 01-28-2011 & LRIS & 0.7 & 1.62 & 0.6 & 3600 & 0.419 & 2.95 & $281 \pm 22$ & 33 & $120$ \\ 
SL2SJ085826-014300 & 11-30-2011 & LRIS & 1.0 & 1.62 & 0.9 & 3600 & 0.580 & $\cdots$ & $233 \pm 25$ & $\cdots$ & $160$ \\ 
SL2SJ090106-025906 & 01-07-2011 & XSHOOTER & 0.9 & 1.60 & 0.7 & 2800 & 0.670 & 1.19 & $\cdots$ & 7 & $49$ \\ 
SL2SJ090407-005952 & 12-30-2010 & XSHOOTER & 0.9 & 1.60 & 0.7 & 2800 & 0.611 & 2.36 & $183 \pm 21$ & 22 & $52$ \\ 
SL2SJ095921+020638 & 02-02-2011 & XSHOOTER & 0.9 & 1.60 & 0.7 & 2800 & 0.552 & 3.35 & $188 \pm 22$ & 17 & $47$ \\ 
SL2SJ135847+545913 & 04-29-2011 & LRIS & 1.0 & 1.62 & 0.8 & 2700 & 0.510 & $\cdots$ & $287 \pm 22$ & 28 & $150$ \\ 
  & 03-22-2013 & GNIRS & 0.675 & $\cdots$ & 0.7 & 7200 &   &   & $\cdots$ & $\cdots$ & $\cdots$ \\ 
SL2SJ135949+553550 & 03-17-2010 & LRIS & 1.0 & 1.62 & 0.7 & 5400 & 0.783 & 2.77 & $228 \pm 29$ & 9 & $150$ \\ 
  & 04-29-2011 & LRIS & 1.0 & 1.62 & 0.9 & 5400 &   &   & $234 \pm 28$ & 12 & $150$ \\ 
SL2SJ140123+555705 & 07-20-2006 & LRIS & 1.5 & 3.36 & 0.8 & 1200 & 0.527 & $\cdots$ & $332 \pm 25$ & 10 & $210$ \\ 
SL2SJ140156+554446 & 04-29-2011 & LRIS & 1.0 & 1.62 & 0.8 & 2700 & 0.464 & $\cdots$ & $297 \pm 22$ & 34 & $150$ \\ 
SL2SJ140221+550534 & xx-xx-2xxx & SDSS & $\cdots$ & $\cdots$ & $\cdots$ & $\cdots$ & 0.412 & $\cdots$ & $\cdots$ & $\cdots$ & $\cdots$ \\ 

\enddata
\tablecomments{Summary of spectroscopic observations and derived parameters.}
\end{deluxetable*}
% %%%%%%%%%%%%%%%%%%%%%%%%%%%%%%%%%%%%%

% %%%%%%%%%%%%%%%%%%%%%%%%%%%%%%%%%%%%%
\renewcommand{\arraystretch}{1.10} 
\begin{deluxetable*}{lccccccccccc}
%\tablenum{\ref{table:allspec}} %WHY IS THIS NOT WORKING??
\tablenum{2}
\tablewidth{0pt}
\tablecaption{Spectroscopic observations {\it (continued)}.}
\tabletypesize{\scriptsize}
\tablehead{
\colhead{Name} & \colhead{obs. date} & \colhead{Instrument} & 
\colhead{slit} & \colhead{width} & \colhead{seeing} & \colhead{exp. time} & \colhead{$z_d$} & \colhead{$z_s$} 
& \colhead{$\sigma$} & \colhead{S/N} & \colhead{res.} \\
& & & ($''$) & ($''$) & ($''$) & (s) & & & (km/s) & (\AA$^{-1}$) & 
(km/s)
}
\startdata
SL2SJ140454+520024 & 04-30-2011 & LRIS & 1.0 & 1.62 & 0.9 & 1800 & 0.456 & 1.59 & $342 \pm 20$ & 38 & $140$ \\ 
SL2SJ140546+524311 & 04-29-2011 & LRIS & 1.0 & 1.62 & 0.8 & 2700 & 0.526 & 3.01 & $284 \pm 21$ & 30 & $140$ \\ 
  & 03-26-2013 & GNIRS & 0.675 & $\cdots$ & 0.5 & 4800 &   &   & $\cdots$ & $\cdots$ & $\cdots$ \\ 
SL2SJ140614+520253 & 07-20-2006 & LRIS & 1.5 & 3.36 & 0.8 & 1200 & 0.480 & $\cdots$ & $247 \pm 29$ & 11 & $190$ \\ 
SL2SJ140650+522619 & 04-29-2011 & LRIS & 1.0 & 1.62 & 0.9 & 3600 & 0.716 & 1.47 & $253 \pm 19$ & 15 & $150$ \\ 
  & 04-30-2011 & LRIS & 1.0 & 1.62 & 0.9 & 3600 &   &   & $247 \pm 20$ & 16 & $160$ \\ 
SL2SJ141137+565119 & 01-14-2010 & LRIS & 1.0 & 1.62 & 1.3 & 2700 & 0.322 & 1.42 & $214 \pm 23$ & 35 & $470$ \\ 
SL2SJ142003+523137 & 04-30-2011 & LRIS & 1.0 & 1.62 & 0.9 & 2700 & 0.354 & 1.41 & $\cdots$ & 4 & $150$ \\ 
SL2SJ142031+525822 & 04-29-2011 & LRIS & 1.0 & 1.62 & 0.8 & 1800 & 0.380 & 0.99 & $246 \pm 23$ & 24 & $150$ \\ 
SL2SJ142059+563007 & 04-29-2011 & LRIS & 1.0 & 1.62 & 0.9 & 1800 & 0.483 & 3.12 & $\cdots$ & 20 & $\cdots$ \\ 
  & 04-30-2011 & LRIS & 1.0 & 1.62 & 0.8 & 1800 &   &   & $228 \pm 19$ & 18 & $160$ \\ 
SL2SJ142731+551645 & 04-30-2011 & LRIS & 1.0 & 1.62 & 0.8 & 3600 & 0.511 & 2.58 & $\cdots$ & 12 & $150$ \\ 
SL2SJ220329+020518 & 08-06-2010 & LRIS & 1.0 & 1.62 & 0.9 & 2700 & 0.400 & 2.15 & $213 \pm 21$ & 36 & $170$ \\ 
SL2SJ220506+014703 & 10-06-2010 & XSHOOTER & 0.9 & 1.60 & 0.8 & 2800 & 0.476 & 2.53 & $317 \pm 30$ & 29 & $49$ \\ 
SL2SJ220629+005728 & 09-13-2007 & LRIS & 1.0 & 1.68 & 0.7 & 2700 & 0.704 & $\cdots$ & $290 \pm 39$ & 6 & $230$ \\ 
SL2SJ221326-000946 & 09-09-2009 & LRIS & 1.0 & 1.62 & 1.0 & 1800 & 0.338 & 3.45 & $165 \pm 20$ & 30 & $150$ \\ 
  & 07-29-2011 & XSHOOTER & 0.9 & 1.60 & 0.8 & 2800 &   &   & $177 \pm 21$ & 32 & $56$ \\ 
SL2SJ221407-180712 & 09-13-2007 & LRIS & 1.0 & 1.68 & 0.7 & 2700 & 0.651 & $\cdots$ & $200 \pm 24$ & 6 & $220$ \\ 
SL2SJ221852+014038 & 08-06-2010 & LRIS & 1.0 & 1.62 & 0.9 & 2700 & 0.564 & $\cdots$ & $305 \pm 23$ & 28 & $170$ \\ 
  & 11-10-2012 & GNIRS & 0.675 & $\cdots$ & 0.7 & 3600 &   &   & $\cdots$ & $\cdots$ & $\cdots$ \\ 
SL2SJ221929-001743 & 09-14-2007 & LRIS & 0.7 & 1.68 & 0.6 & 1800 & 0.289 & 1.02 & $189 \pm 20$ & 23 & $420$ \\ 
SL2SJ222012+010606 & 08-18-2012 & DEIMOS & 1.0 & 1.88 & 1.2 & 3600 & 0.232 & 1.07 & $127 \pm 15$ & 14 & $170$ \\ 
SL2SJ222148+011542 & 11-11-2012 & GNIRS & 0.675 & $\cdots$ & 0.7 & 3600 & 0.325 & 2.35 & $\cdots$ & $\cdots$ & $\cdots$ \\ 
  & 08-18-2012 & DEIMOS & 1.0 & 1.88 & 1.2 & 3600 &   &   & $222 \pm 23$ & 25 & $160$ \\ 
  & 10-01-2012 & XSHOOTER & 0.9 & 1.60 & 1.0 & 1400 &   &   & $\cdots$ & $\cdots$ & $\cdots$ \\ 
SL2SJ222217+001202 & 08-06-2010 & LRIS & 1.0 & 1.62 & 0.9 & 900 & 0.436 & 1.36 & $221 \pm 22$ & 13 & $170$ \\ 
  & 11-01-2010 & LRIS & 1.0 & 1.62 & 0.9 & 900 &   &   & $200 \pm 29$ & 10 & $150$ \\ 

\enddata
\tablecomments{Summary of spectroscopic observations and derived parameters.}
\end{deluxetable*}
% %%%%%%%%%%%%%%%%%%%%%%%%%%%%%%%%%%%%%

% %%%%%%%%%%%%%%%%%%%%%%%%%%%%%%%%%%%%%%%
% MASSIVE MULTI-PAGE FIGURE OF SPECTRA
% %%%%%%%%%%%%%%%%%%%%%%%%%%%%%%%%%%%%%%%
%\input{figs/rx/specfig.tex}
\begin{figure*}
\centering\includegraphics[width=0.9\textwidth]{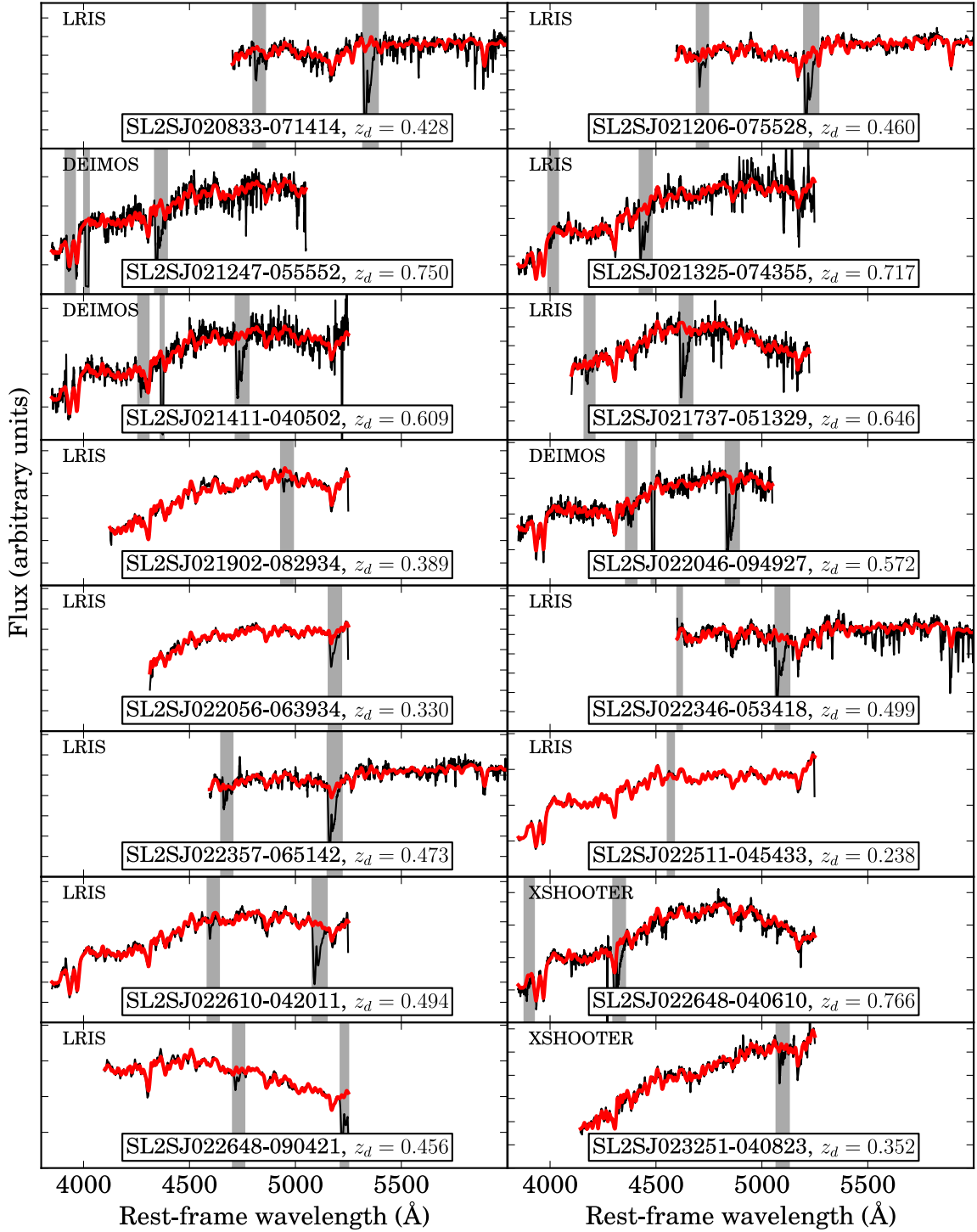}\medskip
\caption{1d spectra of SL2S lenses and lens candidates (in black). Where available, we overplot the best fit spectrum obtained for the velocity dispersion fitting (in red). Only the rest-frame wavelength region used in the fit is shown. Vertical gray bands are regions of the spectrum masked out of the fit and typically correspond to atmospheric absorption features. Each plot indicates the redshift of the galaxy and the instrument used to acquire the data shown.}
\label{fig:spec}
\end{figure*}
\begin{figure*}
\figurenum{\ref{fig:spec}}
\centering\includegraphics[width=0.9\textwidth]{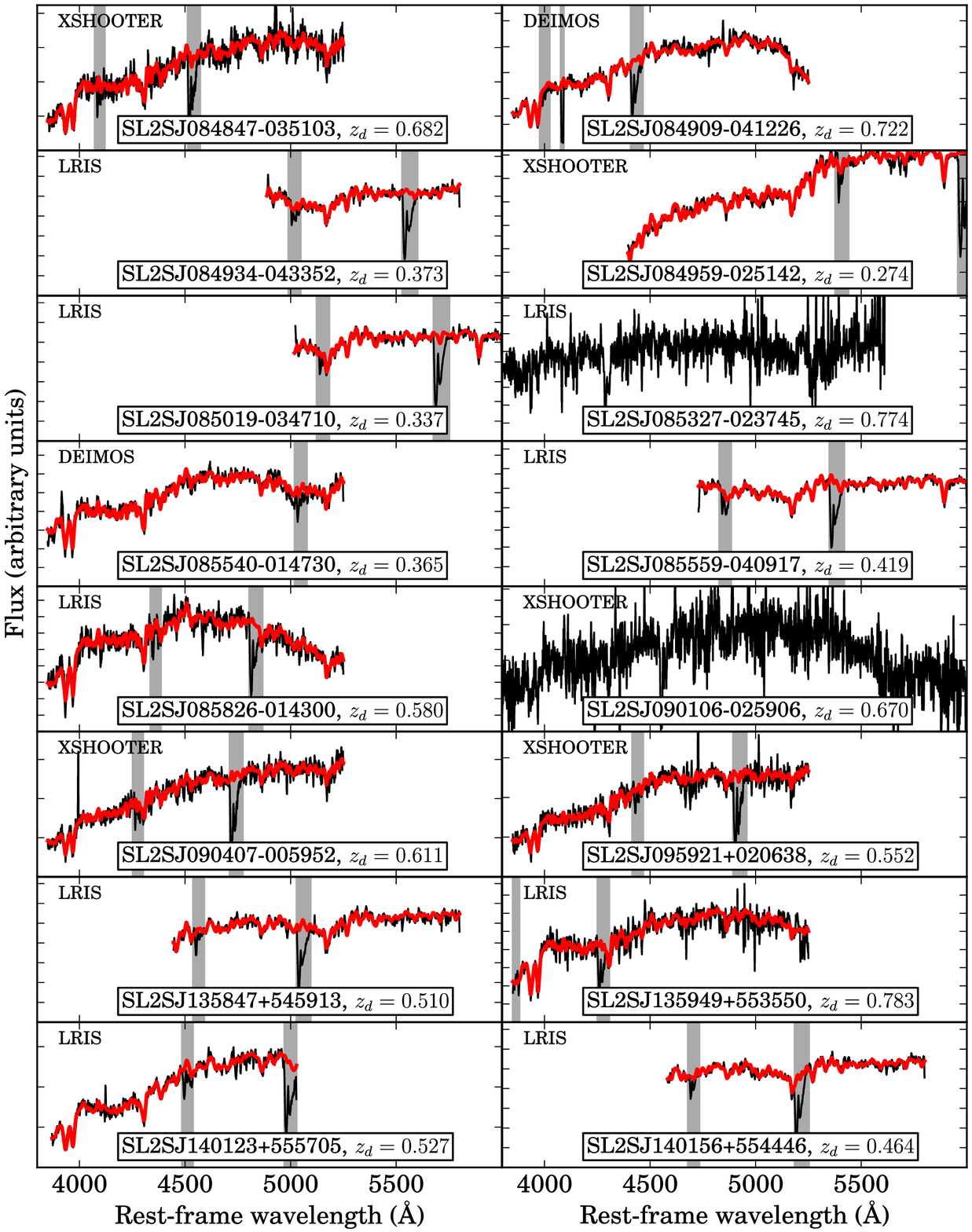}\medskip
\caption{{\it continued.}}
\end{figure*}
\begin{figure*}
\figurenum{\ref{fig:spec}}
\centering\includegraphics[width=0.9\textwidth]{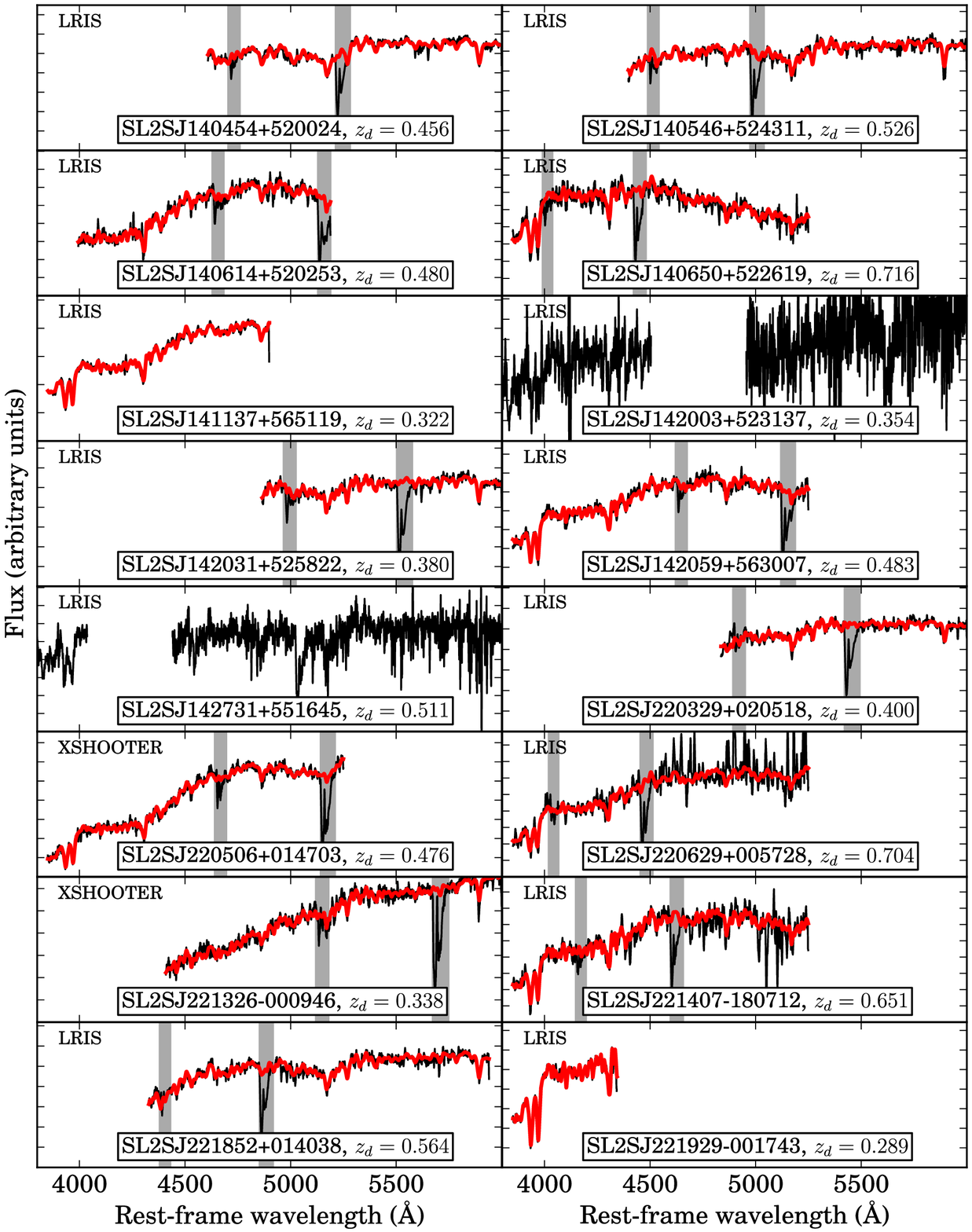}\medskip
\caption{{\it continued.}}
\end{figure*}
\begin{figure*}
\figurenum{\ref{fig:spec}}
\centering\includegraphics[width=0.9\textwidth]{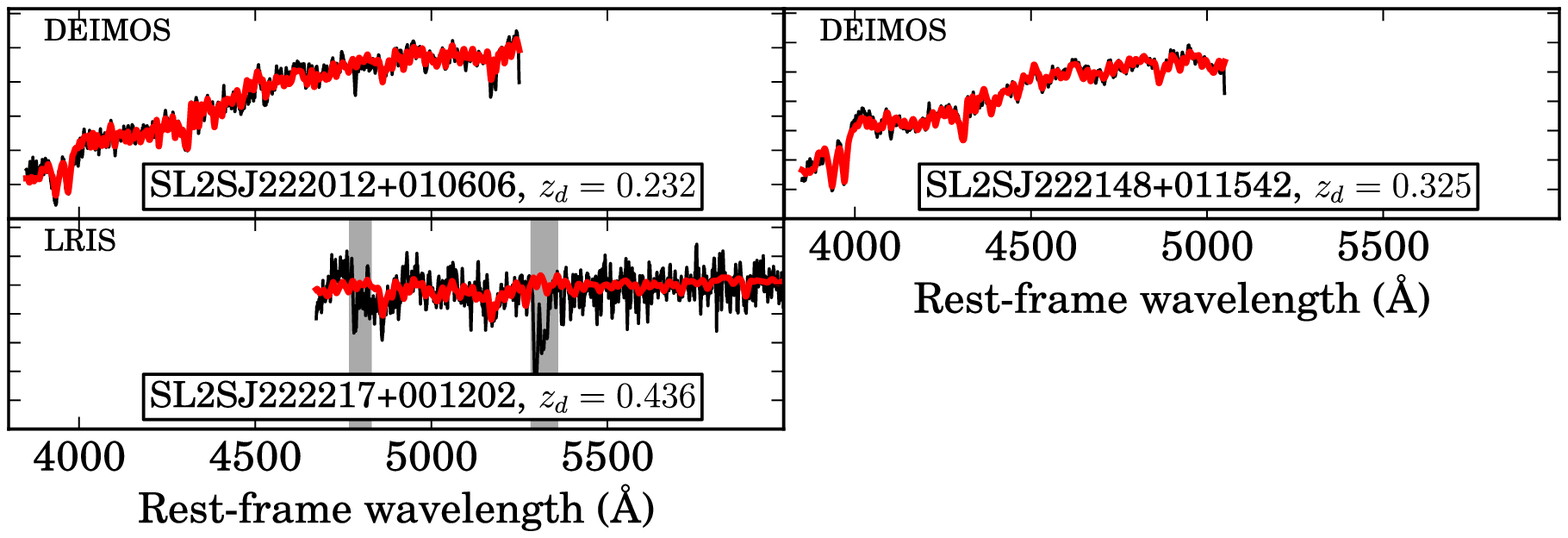}\medskip
\caption{{\it continued.}}
\end{figure*}
% %%%%%%%%%%%%%%%%%%%%%%%%%%%%%%%%%%%%%%%

% %%%%%%%%%%%%%%%%%%%%%%%%%%%%%%%%%%%%%%%
% Source continuum spectra figure:
% %%%%%%%%%%%%%%%%%%%%%%%%%%%%%%%%%%%%%%%

% %%%%%%%%%%%%%%%%%%%%%%%%%%%%%%%%%%%%%%%
% %%%%%%%%%%%%%%%%%%%%%%%%%%%%%%%%%%%%%%%

%-------------------------------------------------------------------------------

\section{Sample characterization}\label{sect:class}

\begin{figure}
\includegraphics[width=\columnwidth]{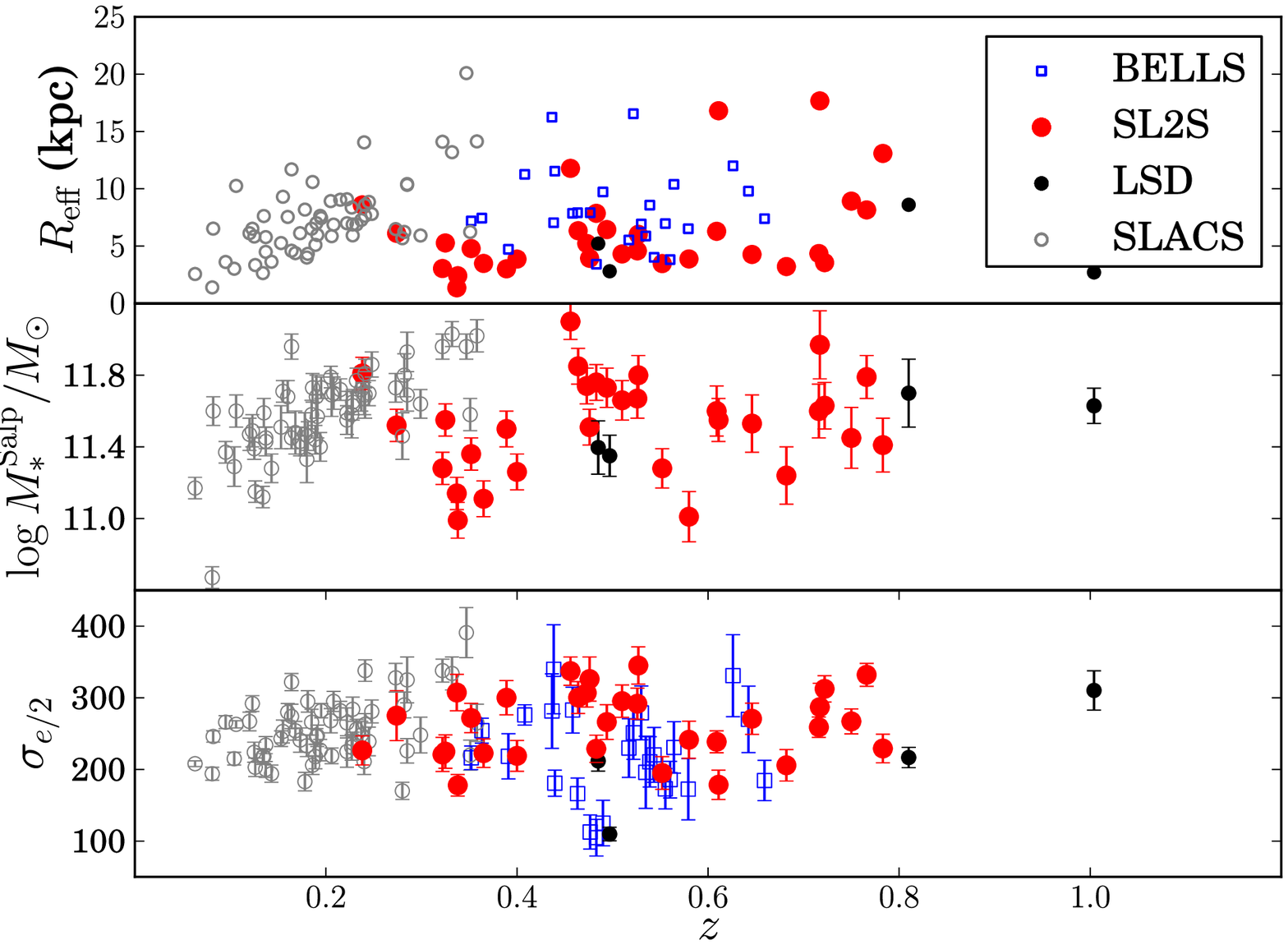}
\caption{\label{fig:classevol} Effective radius, stellar mass and velocity dispersion of lenses as a function of redshift. %The solid line shows the best fit linear relation $y = az + b$ to the SL2S lenses only.}
}
\end{figure}

In Paper III we presented effective radii, magnitudes, stellar masses
and Einstein radii of our lenses. Here we complement this information
with lens and source redshifts, and lens velocity dispersions. It is
possible at this point to look at the distribution of our lenses in the
parameter space defined by these quantities. Since our scientific goal
is to measure the evolution in the mean
density slope with time, it is very
important to assess whether other observables appear to evolve in our
sample. In \Fref{fig:classevol} we plot the effective radii, stellar masses
and velocity dispersions as a function of redshift for all our objects,
and also for lenses from other surveys. 
Throughout this paper, when dealing
with stellar masses we refer to values measured from stellar population
synthesis fitting based on a Salpeter initial mass function (IMF). For a
fair comparison, all velocity dispersions, which are measured within
rectangular apertures of arbitrary sizes, are transformed into velocity
dispersions within a circular aperture, $\sigmae2$, with radius
$R_{\mathrm{eff}}/2$ following the prescription of
\citet{jorgensen1995}. The values of $\sigmae2$ for individual SL2S
lenses are reported in \Tref{table:lensing}. 

SL2S lenses do not appear
to differ from objects from independent lensing surveys in the average
values of $R_{\mathrm{eff}}$, $M_*$ and $\sigmae2$. As far as trends
with redshift within the SL2S sample are concerned, there is a mild
increase of the stellar mass with $z$ that will need to be taken into
account when discussing the evolution of the mass profile of these
objects. 

\begin{figure}
\includegraphics[width=\columnwidth]{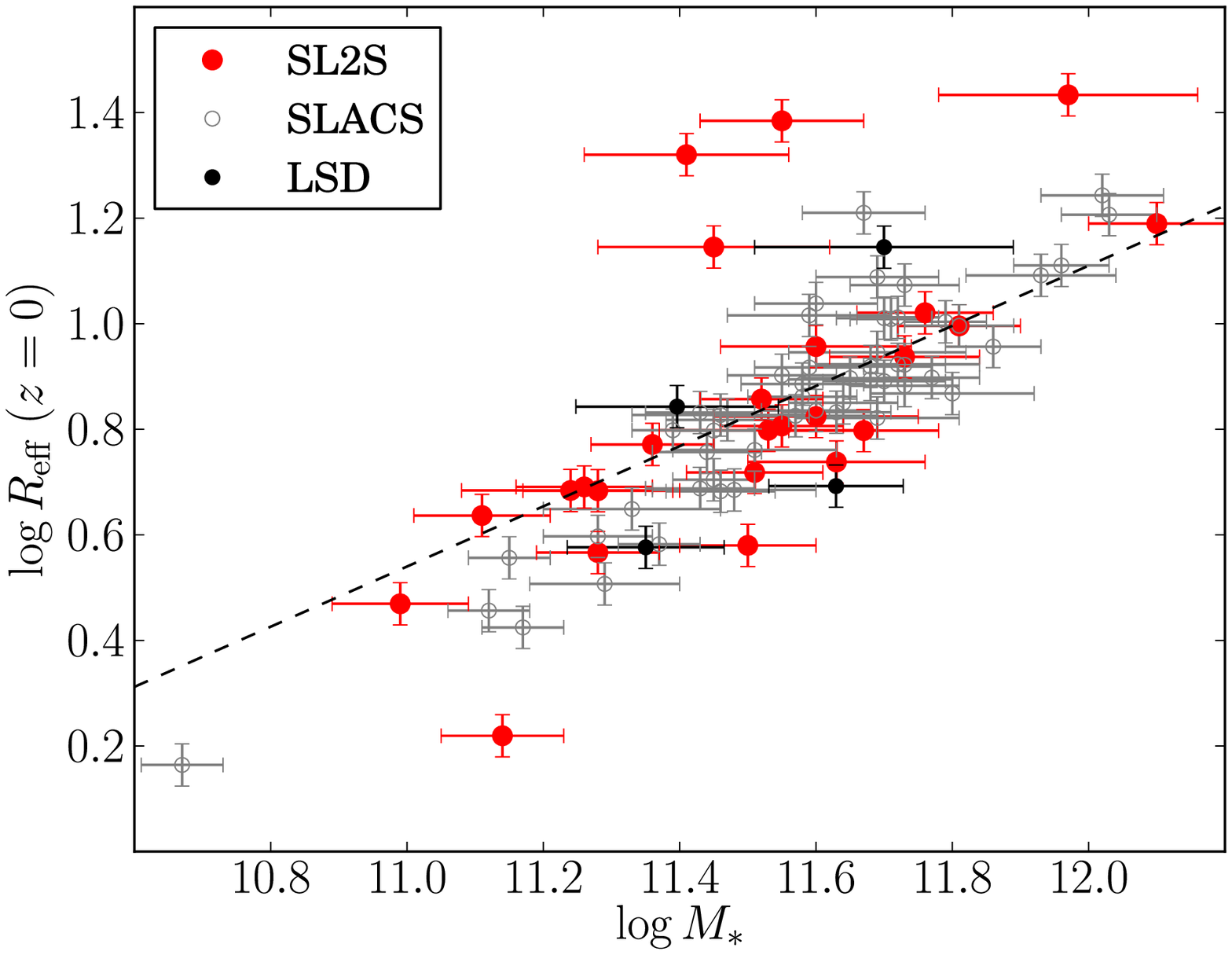}
\caption{\label{fig:mreffz0} Effective radius vs. stellar mass, where $\reff$ values have been corrected for the evolution in the mass-size relation measured by \citet{New++12}: $\log{\reff} (z=0) = \log{\reff} + 0.26z$. The dashed line indicates the mass-radius relation for SDSS galaxies measured by \citet{New++12}.}
\end{figure}

As an additional test, we examine the correlation between mass and effective radius for SL2S, SLACS and LSD lenses and check it against non-lens galaxies. 
The goal is to make sure that these surveys do not preferentially select lenses with a larger or smaller size than typical ETGs of their mass.
The mass-radius relation is seen to evolve with time \citep[e.g.][]{Dam++11,New++12,Cim++12}. We correct for this evolution by considering effective radii evolved to $z=0$ assuming the trend measured by \citet{New++12}: $\log{\reff} (z=0) = \log{\reff} + 0.26z$. 
Effective radii defined in this way are plotted against measured stellar masses in \Fref{fig:mreffz0}, together with the mass-radius relation measured by \citet{New++12} for low-redshift SDSS galaxies.
Points in the plot of \Fref{fig:mreffz0} should not be considered as evolutionary tracks of individual objects, as galaxies grow in mass as well as in size.
For a given object, its redshift-evolved size $\reff (z=0)$ is equivalent to its measured effective radius rescaled by the average size of galaxies at its redshift and at a reference mass.
This allows us to promptly display in a single plot how our lenses compare, in terms of size, to other galaxies of the same mass, regardless of redshift.
%$\reff (z=0)$ values are effective radii rescaled in terms of the average size at given redshift, and are only introduced in order to promptly compare 
%These redshift-evolved sizes should not be considered as evolutionary tracks of individual objects, as galaxies grow in mass as well as in effective radius. $\reff (z=0)$ values
%are only defined in order to reduce the scatter introduced by comparing masses and sizes for objects at different cosmic times, when the mean effective radius for objects of a given mass is different.
%Effective radii defined in this way are plotted against measured stellar masses in \Fref{fig:mreffz0}, together with the mass-radius relation measured by \citet{New++12} for low-redshift SDSS galaxies.
We see from \Fref{fig:mreffz0} that lenses from all surveys lie nicely around the relation found for non-lenses, indicating that our sample of lenses does not appear special when compared to the more general population of galaxies of their redshift.

%The higher redshift objects are also seen to have larger
%effective radii. This could be due to pure statistical fluctuations, a
%selection effect, or, since more massive galaxies have on average larger
%sizes, could just reflect the trend with stellar mass. To investigate
%this further we plot in \Fref{fig:reffevol} mass-normalized radii as a
%function of redshift. The mass-normalized radius is defined as the
%effective radius rescaled to a stellar mass of $10^{11}M_\odot$ assuming
%a mass-radius relation $R_{\mathrm{eff}} \propto M_*^\alpha$. Here we
%use $\alpha = 0.55$ following \citet{Cim++12} and compare the trend in
%the mass-normalized radius with the evolution observed on a much larger
%sample of early-type galaxies by \citet{Cim++12}. Some of the highest
%redshift lenses appear to have larger radii than the average early-type
%galaxy of their mass. This also needs to be taken into account when
%discussing trends on quantities derived from $R_{\mathrm{eff}}$.

% %%%%%%%%%%%%%%%%%%%%%%%%%%%%%%%%%%%%%%%
% %%%%%%%%%%%%%%%%%%%%%%%%%%%%%%%%%%%%%%%

% %%%%%%%%%%%%%%%%%%%%%%%%%%%%%%%%%%%%%%%
%\begin{figure}
%\includegraphics[width=\columnwidth]{figs/rx/massradius.eps}
%\caption{\label{fig:reffevol} Mass-normalized radius, defined following \citet{Cim++12} as the effective radius divided by $M_{11}^{\alpha}$, where $M_{11}$ is the stellar mass in units of $10^{11}M_\odot$, as a function of redshift. The dotted line is the mean evolution with redshift of the mass-normalized radius found by \citet{Cim++12}}
%\end{figure}

% %%%%%%%%%%%%%%%%%%%%%%%%%%%%%%%%%%%%%%%

%-------------------------------------------------------------------------------

\section{Power law models}\label{sect:gammap}

We now proceed to combine lensing measurements with stellar kinematics
information to infer the total mass density profile of each lens
galaxy.  We follow the now standard procedure in lensing and dynamics
studies \citep{T+K02a}, as used by \citet{Ruf++11}.  We model the total
(dark matter + stars) mass profile as a spherical power law $\rho(r)
\propto r^{-\gamma'}$ 
% SHS: added this phrase:
in the kinematic analysis. 
The free parameters of the model are the slope
$\gamma'$, and the mass normalization.  For a given model we calculate
the line of sight velocity dispersion within the rectangular aperture
of our observation, broadened by the seeing, through the spherical
Jeans equation. We assume isotropic orbits and a de Vaucouleurs
profile for the distribution of tracers \citep{deV48}, with effective radius fixed
to the observed one.  We then compare the model to the observed
velocity dispersion and Einstein radius to derive posterior
probability densities for the free parameters. In spite of the clear
approximations, the method has been shown to be very robust when
compared to results of more sophisticated models \citep[e.g. ][]{Bar++11a}.

The data required for this inference are the Einstein radius of the
lens, the redshift of both the deflector galaxy and the lensed source,
and the velocity dispersion of the lens.  Of the 39 grade A lenses of
the SL2S sample, 25 have all the required data.  For the few systems
with two or more independent measurements of the velocity dispersion,
we use the weighted average.  The inferred values of $\gamma'$ are
reported in Table~\ref{table:lensing}. 

%and plotted as a function of
%redshift, stellar mass and stellar mass density $\Sigma_* = M_*/(2\pi
%R_{\mathrm{eff}}^2)$ in Figures
%\ref{fig:gammaprime},
%\ref{fig:gammap_mstar} and \ref{fig:gammap_sstar} respectively.

% %%%%%%%%%%%%%%%%%%%%%%%%%%%%%%%%%%%%%%%
\renewcommand{\arraystretch}{1.10} 
\begin{deluxetable*}{lccccccc}
\tablewidth{0pt}
\tabletypesize{\small}
\tablecaption{Lensing and dynamics.}
\tabletypesize{\footnotesize}
\tablehead{
\colhead{Name} & $z_d$ & $R_{\mathrm{eff}}$ & $R_{\mathrm{Ein}}$  
& \colhead{$\sigmae2$} & $\log{M_*^{\mathrm{Salp}}/M_\odot}$ & $\gamma'$ & Notes \\
 & & (kpc) & (kpc) & (km s$^{-1}$) & & & 
}

\startdata
SL2SJ021247-055552 & $0.750$ & $ 8.92$ & $ 9.33$ & $267\pm17$ & $11.45\pm0.17$ & $2.05\pm0.09$ &  \\ 
SL2SJ021325-074355 & $0.717$ & $17.67$ & $17.22$ & $287\pm33$ & $11.97\pm0.19$ & $1.79\pm0.12$ &  \\ 
SL2SJ021411-040502 & $0.609$ & $ 6.29$ & $ 9.48$ & $238\pm15$ & $11.60\pm0.14$ & $1.85\pm0.07$ &  \\ 
SL2SJ021737-051329 & $0.646$ & $ 4.27$ & $ 8.80$ & $270\pm21$ & $11.53\pm0.16$ & $2.02\pm0.09$ &  \\ 
SL2SJ021902-082934 & $0.389$ & $ 3.01$ & $ 6.88$ & $300\pm23$ & $11.50\pm0.10$ & $2.26\pm0.08$ &  \\ 
SL2SJ022511-045433 & $0.238$ & $ 8.59$ & $ 6.65$ & $226\pm20$ & $11.81\pm0.09$ & $1.78\pm0.10$ &  \\ 
SL2SJ022610-042011 & $0.494$ & $ 6.44$ & $ 7.23$ & $266\pm24$ & $11.73\pm0.11$ & $2.01\pm0.12$ &  \\ 
SL2SJ023251-040823 & $0.352$ & $ 4.78$ & $ 5.15$ & $271\pm20$ & $11.36\pm0.09$ & $2.39\pm0.10$ &  \\ 
SL2SJ084847-035103 & $0.682$ & $ 3.21$ & $ 6.02$ & $205\pm21$ & $11.24\pm0.16$ & $1.84\pm0.13$ &  \\ 
SL2SJ084909-041226 & $0.722$ & $ 3.55$ & $ 7.94$ & $312\pm18$ & $11.63\pm0.13$ & $2.14\pm0.07$ &  \\ 
SL2SJ084959-025142 & $0.274$ & $ 6.11$ & $ 4.84$ & $275\pm34$ & $11.52\pm0.09$ & $2.33\pm0.17$ &  \\ 
SL2SJ085019-034710 & $0.337$ & $ 1.35$ & $ 4.48$ & $307\pm25$ & $11.14\pm0.09$ & $2.45\pm0.07$ & disky \\ 
SL2SJ085540-014730 & $0.365$ & $ 3.48$ & $ 5.21$ & $222\pm19$ & $11.11\pm0.10$ & $2.15\pm0.11$ &  \\ 
SL2SJ090407-005952 & $0.611$ & $16.81$ & $ 9.47$ & $178\pm20$ & $11.55\pm0.12$ & $1.48\pm0.11$ &  \\ 
SL2SJ095921+020638 & $0.552$ & $ 3.47$ & $ 4.73$ & $195\pm22$ & $11.28\pm0.11$ & $2.12\pm0.16$ &  \\ 
SL2SJ135949+553550 & $0.783$ & $13.08$ & $ 8.52$ & $229\pm19$ & $11.41\pm0.15$ & $1.86\pm0.14$ &  \\ 
SL2SJ140454+520024 & $0.456$ & $11.78$ & $14.80$ & $337\pm19$ & $12.10\pm0.10$ & $1.95\pm0.06$ &  \\ 
SL2SJ140546+524311 & $0.526$ & $ 4.58$ & $ 9.48$ & $291\pm21$ & $11.67\pm0.11$ & $2.14\pm0.08$ &  \\ 
SL2SJ140650+522619 & $0.716$ & $ 4.35$ & $ 6.79$ & $258\pm14$ & $11.60\pm0.15$ & $2.01\pm0.07$ &  \\ 
SL2SJ141137+565119 & $0.322$ & $ 3.04$ & $ 4.34$ & $220\pm23$ & $11.28\pm0.09$ & $2.15\pm0.15$ &  \\ 
SL2SJ142059+563007 & $0.483$ & $ 7.86$ & $ 8.39$ & $228\pm19$ & $11.76\pm0.10$ & $1.93\pm0.11$ &  \\ 
SL2SJ220329+020518 & $0.400$ & $ 3.86$ & $10.49$ & $218\pm21$ & $11.26\pm0.10$ & $1.77\pm0.09$ &  \\ 
SL2SJ220506+014703 & $0.476$ & $ 3.93$ & $ 9.87$ & $326\pm30$ & $11.51\pm0.10$ & $2.19\pm0.09$ &  \\ 
SL2SJ221326-000946 & $0.338$ & $ 2.41$ & $ 5.17$ & $177\pm15$ & $10.99\pm0.10$ & $1.89\pm0.09$ & disky \\ 
SL2SJ222148+011542 & $0.325$ & $ 5.27$ & $ 6.59$ & $224\pm23$ & $11.55\pm0.09$ & $1.96\pm0.13$ &  \\ 

\enddata
\tablecomments{\label{table:lensing}
Summary of lensing and dynamics measurements.
}
\end{deluxetable*}
% %%%%%%%%%%%%%%%%%%%%%%%%%%%%%%%%%%%%%%%

% - - - - - - - - - - - - - - - - - - - - - - - - - - - - - - - - - - -

\subsection{The meaning of $\gamma'$}

Before analyzing the measurements in a statistical sense we need to
understand what physical properties the quantity $\gamma'$ is most
sensitive to.  Observations \citep{Son++12} and simple arguments
(galaxies have a finite mass) suggest that the true density profile
deviates from a pure power law, 
% SHS: added this phrase:
particularly at large radii.  
Thus our power law fits to the
lensing and kinematics data must be interpreted as an approximation of
the average density slope over a radial range explored by our data.
Since for a typical lens both the Einstein radius and the velocity
dispersion probe the region within the effective radius, we expect
that the inferred $\gamma'$ will be close to the mean density slope
within $R_{\mathrm{eff}}$, as suggested by \citet{D+T13}.  
%\QUERY{PJM}{Perhaps this statement could be motivated by a reference to
%\citet{D+T13} -- before I read further down, I was wondering 
%whether you meant ``at''
%instead of ``within''\ldots}

However we would like to be more quantitative and explore the two
following questions: what kind of average over the true density
profile $\rho(r)$ best reproduces the lensing+dynamics $\gamma'$? How
sensitive to the ratio $\rein/\reff$ is the measured $\gamma'$ for a
fixed galaxy mass profile?  The former issue is relevant when
comparing theoretical models to lensing and dynamics measurements.
The latter is important when trying to measure trends of $\gamma'$
with redshift: the ratio $\rein/\reff$ typically increases 
for purely geometrical reasons, and a dependence of $\gamma'$ on
$\rein/\reff$ could in principle bias the inference on the evolution
of the slope.  In order to answer these questions we simulate
$\gamma'$ measurements on a broad range of model mass profiles and
compare these with the true density slopes.  We
consider a pure de Vaucouleurs profile, a sum of a de Vaucouleurs
profile with a Navarro, Frenk \& White \citep{NFW97} profile with two
values of the dark matter mass fraction $f_{\mathrm{DM}}$
within the 3d effective radius, and the most probable total density
profile from the bulge + halo decomposition of the gravitational lens
SDSSJ0946+1006 by \citet{Son++12}.  None of these model profiles is a
pure power law. We emphasize that the range of models is chosen to be
broader than what is likely to be found in real galaxies based on the
detailed analysis of SLACS systems by \cite{Bar++11a}. 

\begin{figure}
\includegraphics[width=\columnwidth]{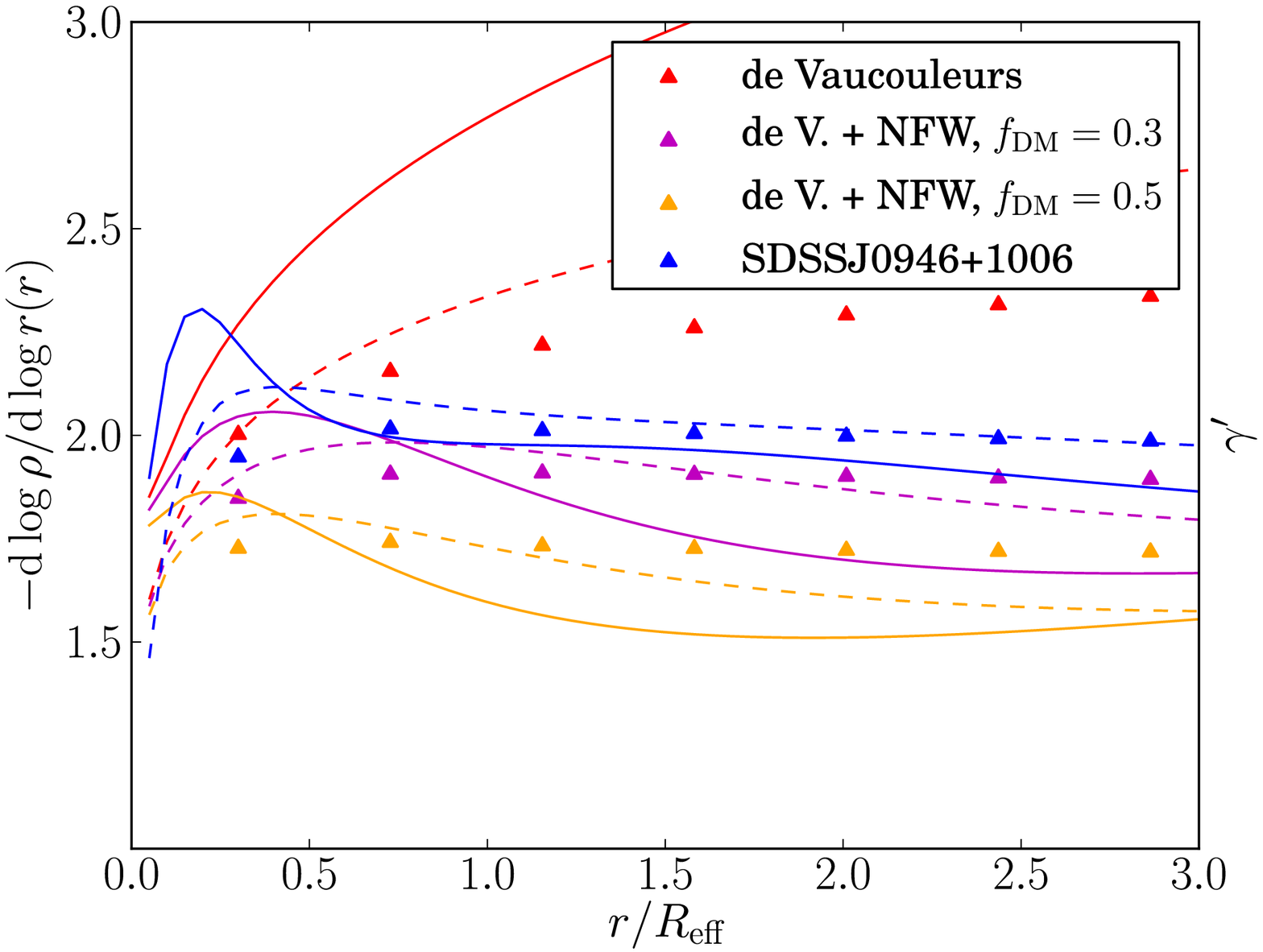}
\caption{\label{fig:model_gammap} {\em Solid lines:} Local logarithmic
density slope as a function of 3d radius, in units of the effective
radius. {\em Dashed lines:} mass-weighted density slope within radius
$r$. {\em Triangles:} lensing+dynamics $\gamma'$ for $\rein=r$.
Different colors indicate the different model mass profiles listed in
the body text.  }
\end{figure}

We again use the spherical Jeans equation to calculate the central velocity
dispersion for each of these model galaxies and then fit power law
density profiles with fixed total projected mass within different
Einstein radii.  These simulated measurements of $\gamma'$ are plotted
in \Fref{fig:model_gammap} as a function of $\rein/\reff$ for each
model profile. In the same plot we show the local logarithmic density
slope $-\mathrm{d}\log{\rho}/\mathrm{d}\log{r}$ as a function of $r$, and also the
{\it mass-weighted density slope within radius $r$}
\begin{equation}
\left<\gamma'(r)\right>_M = \frac{1}{M(r)}\int_0^{r} \gamma'(r')4\pi r^2 \rho(r')dr',
\end{equation}
which has been suggested by \citet{D+T13} to be a good proxy for the
lensing + dynamics $\gamma'$.
%\QUERY{RG}{What is the physical motivation for this equation, and why the dark matter matter only at the denominator? I would suggest plotting $\tilde{\gamma'}= 3 - \frac{\mathrm{d} \log M(r)}{\mathrm{d} \log r}$, as both lensing and dynamics are aperture masses to first order.}

% %%%%%%%%%%%%%%%%%%%%%%%%%%%%%%%%%%%%%%%
% %%%%%%%%%%%%%%%%%%%%%%%%%%%%%%%%%%%%%%%

\Fref{fig:model_gammap} shows that measurements of $\gamma'$ (triangles) 
are remarkably independent of the ratio of the Einstein radius to the
effective radius, for all models.  
This is an important result: it means
that the physical interpretation of $\gamma'$ measurements will be
stable against different lenses having different values of
$\rein/\reff$.  
Excluding the pure de Vaucouleurs
model, which is ruled out on many grounds \citep[mass-follows light
models fail to reproduce lensing and dynamical data, for
example][]{K+T03}, the difference between the mass-weighted slope and
the lensing and dynamics slope is generally smaller than the typical
measurement errors on $\gamma'$ of $\sim0.1$, particularly in the
region $0.5\reff < r < \reff$. 
However the radius at which $\gamma'$ and the
mass-weighted slope are closest is slightly different for different
mass profiles, and so it is 
difficult to interpret $\gamma'$
precisely in terms of a mass-weighted slope within a fixed radius. For
very accurate comparisons with lensing and dynamical data, we
recommend simulating a lensing and dynamics measurement of the models.

%-------------------------------------------------------------------------------

\section{Dependence of the mass density profile slope $\gamma'$ on redshift, stellar mass, and effective radius}
\label{sect:euler}

The main goal of this work is to establish whether, and to what extent,
$\gamma'$ varies with redshift across the population of ETGs.  It
is useful to first study the trends of $\gamma'$ on basic parameters
(\Sref{ssec:euler_qual}) in order to gain insights about the
ingredients that will have to be considered in \Sref{ssec:euler_quant}
to carry out a rigorous statistical analysis.

% - - - - - - - - - - - - - - - - - - - - - - - - - - - - - - - - - - - 

\subsection{Qualitative exploration of the dependency of $\gamma'$ on other parameters}
\label{ssec:euler_qual}

\begin{figure}
\includegraphics[width=\columnwidth]{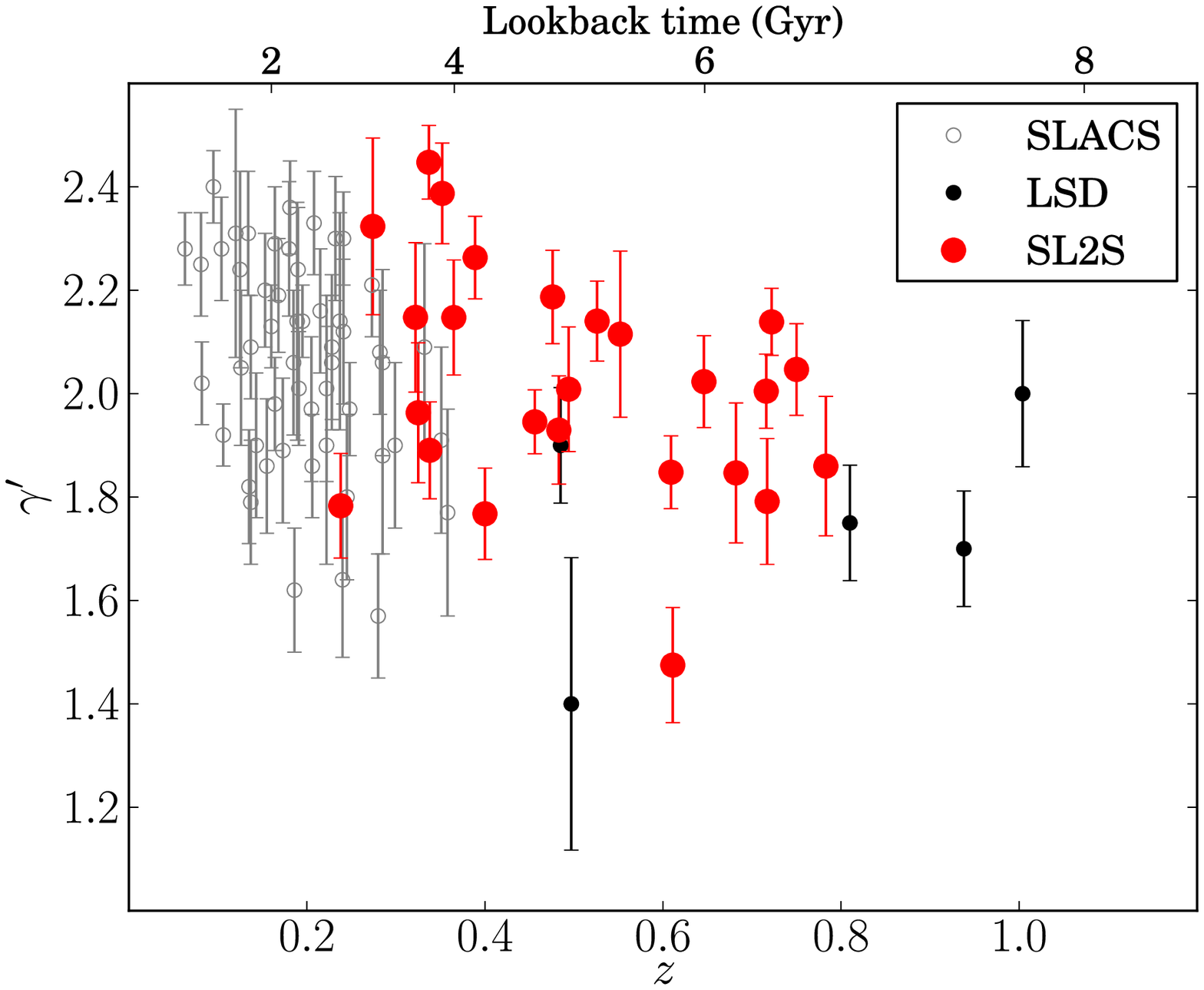}
\caption{\label{fig:gammaprime} Density slope as a function of redshift
for SL2S, SLACS and LSD galaxies.} 
\end{figure}

\begin{figure}
\includegraphics[width=\columnwidth]{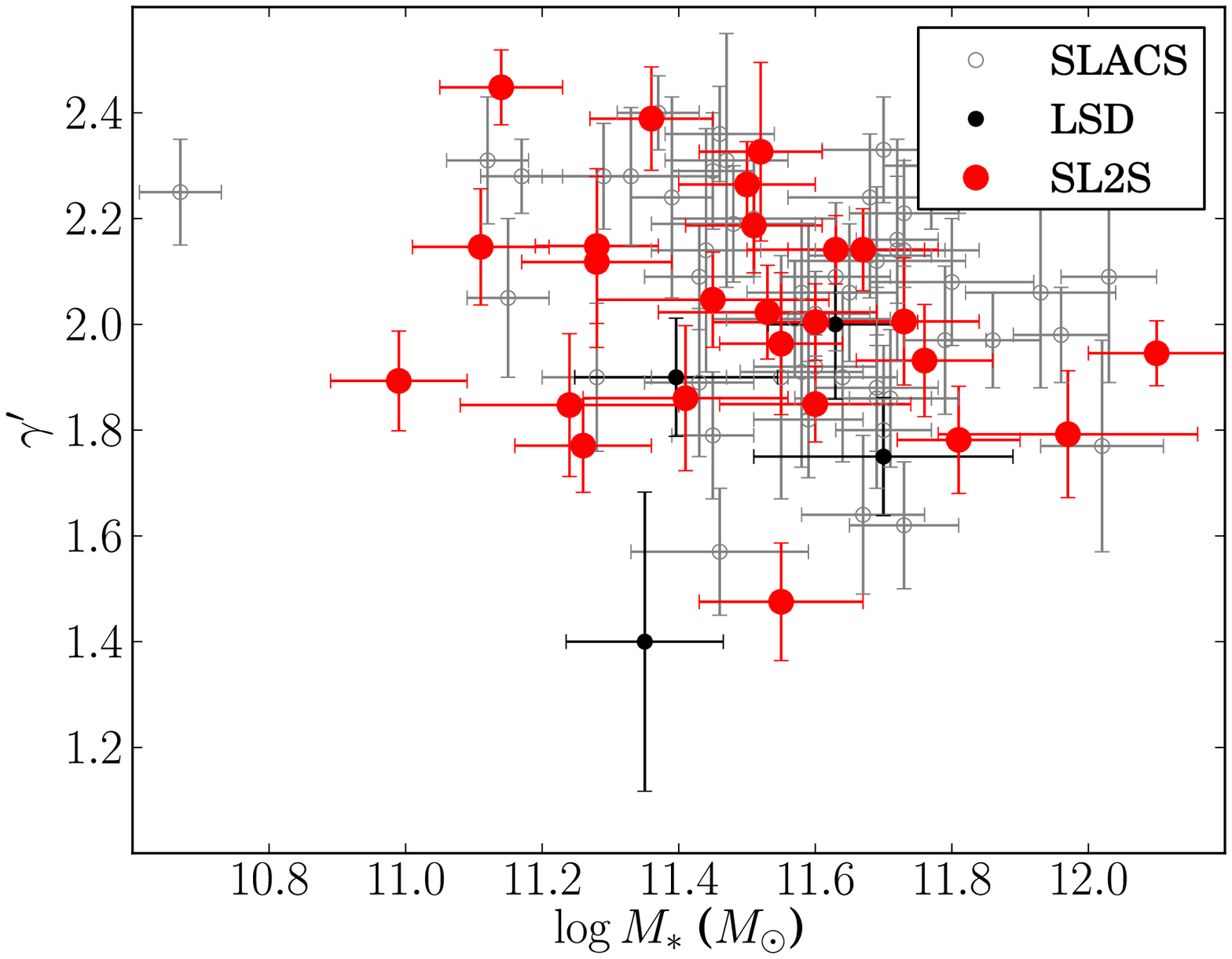}
\caption{\label{fig:gammap_mstar} Density slope as a function of stellar
mass. A Salpeter IMF is assumed.}
\end{figure}

\Fref{fig:gammaprime} shows the individual lens $\gamma'$
values as a function of $z$
for SL2S galaxies, as well as lenses from the SLACS \citep{Aug++10}
and LSD \citep{T+K04} surveys.  A trend of $\gamma'$ with $z$ is
clearly visible, with lower redshift objects having a systematically
steeper slope than higher redshift ones, as previously found by
\citet{Ruf++11} and \citet{Bol++12}.  Before making more quantitative
statements on the time evolution of $\gamma'$ we would like to check
whether the density slope correlates with quantities other than
redshift.  Galaxies grow in mass and size during their evolution, and
a variation of $\gamma'$ with time might be the result of a more
fundamental dependence of the slope on structural properties of ETGs.
Dependences of $\gamma'$ on the effective radius and the stellar
velocity dispersion were explored by \citet{Aug++10}, finding an
anticorrelation with the former and no significant correlation with
the latter.  Here we consider the stellar mass, plotted against
$\gamma'$ in \Fref{fig:gammap_mstar}. A weak trend is visible, with
more massive galaxies having a shallower slope. However the stellar
mass is a rather steep function of redshift in our sample (see
\Fref{fig:classevol}) and the trend seen in \Fref{fig:gammap_mstar}
might just be the result of this selection function.  In fact, if we
fit for a linear dependence of $\gamma'$ on both $z$ and $M_*$ we find
that our data are consistent with $\gamma'$ being independent of $M_*$
at fixed $z$.  

\begin{figure}
\includegraphics[width=\columnwidth]{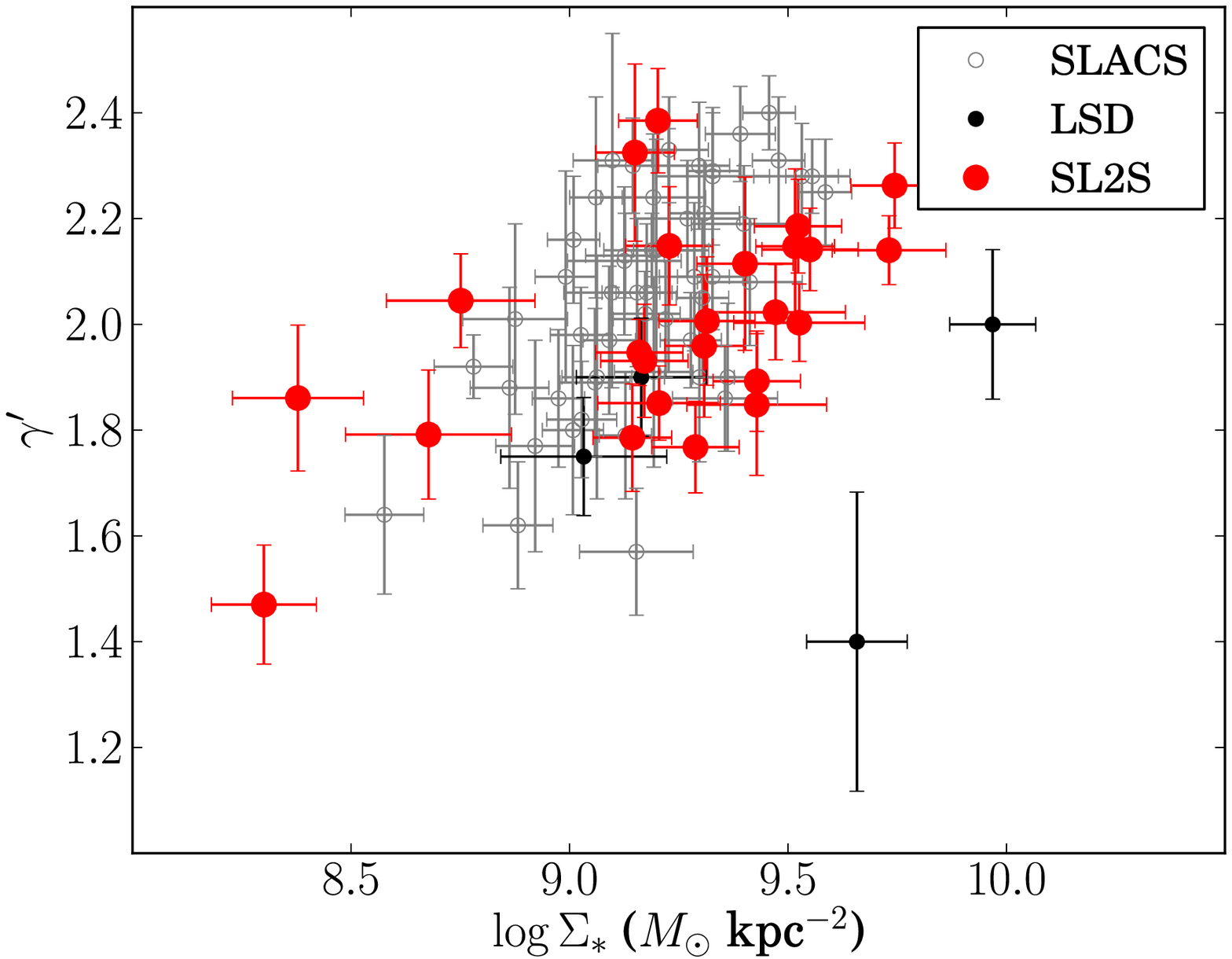}
\caption{\label{fig:gammap_sstar} Density slope as a function of stellar
mass density.}
\end{figure}

A quantity that is expected to correlate with $\gamma'$
is the stellar mass density, $\Sigma_* = M_*/(2\pi \reff^2)$: galaxies
with a more concentrated stellar distribution should have a steeper
overall density profile. This was pointed out by \citet{Aug++10} and \citet{D+T13} and is seen in our data, as shown in
\Fref{fig:gammap_sstar}.  It is therefore important to account for a
dependence of $\gamma'$ on $\Sigma_*$, or on the two independent
variables on which this quantity depends, $\reff$ and $M_*$, when
fitting for the time dependence of the density slope.  This is done in
the next Section.

% %%%%%%%%%%%%%%%%%%%%%%%%%%%%%%%%%%%%%%%
% %%%%%%%%%%%%%%%%%%%%%%%%%%%%%%%%%%%%%%%

% %%%%%%%%%%%%%%%%%%%%%%%%%%%%%%%%%%%%%%%
% %%%%%%%%%%%%%%%%%%%%%%%%%%%%%%%%%%%%%%%

% %%%%%%%%%%%%%%%%%%%%%%%%%%%%%%%%%%%%%%%
% %%%%%%%%%%%%%%%%%%%%%%%%%%%%%%%%%%%%%%%

% - - - - - - - - - - - - - - - - - - - - - - - - - - - - - - - - - - -

\subsection{Quantitative Inference}
\label{ssec:euler_quant}

\begin{figure*}
\includegraphics[width=\textwidth]{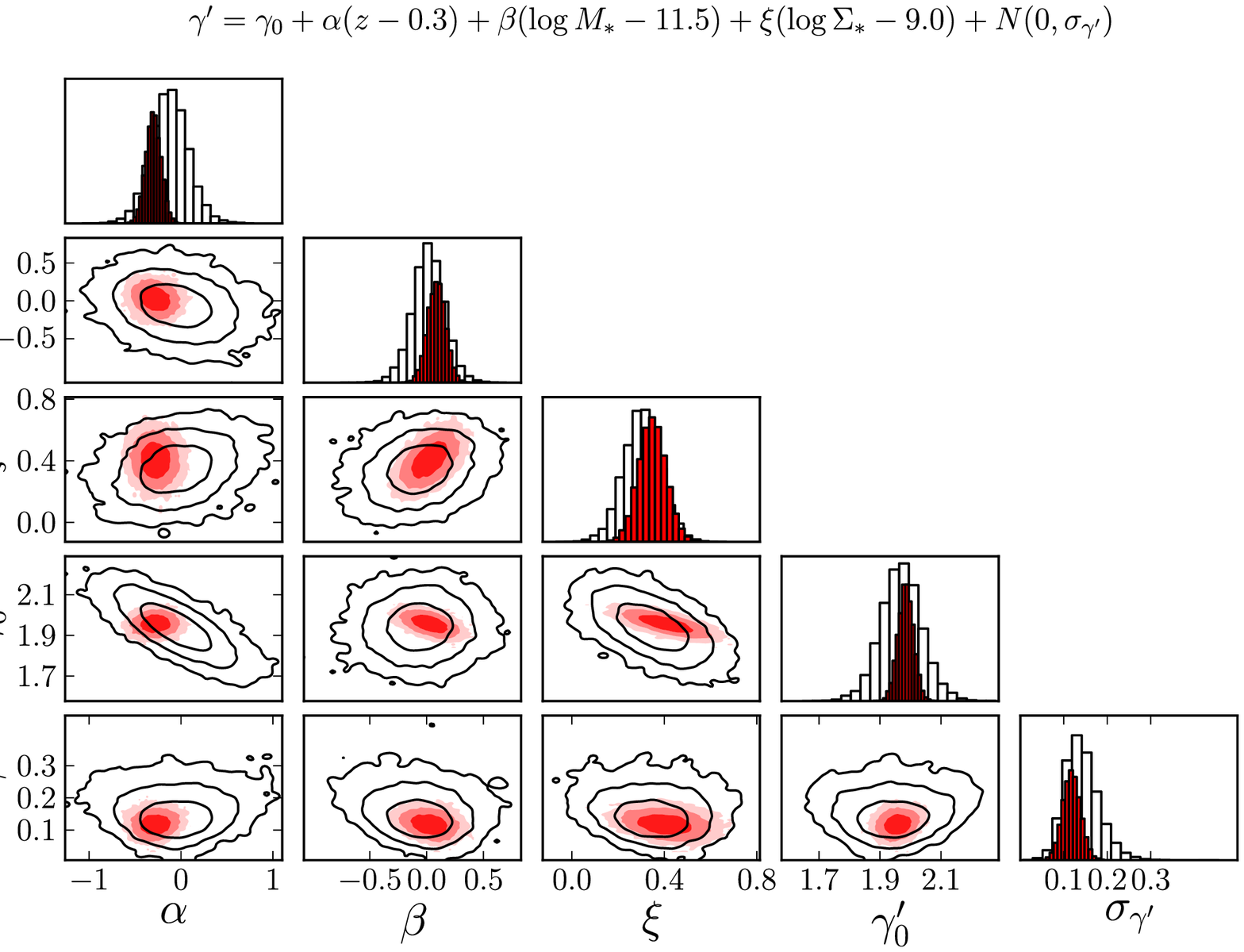}
\caption{\label{fig:gammap_cornerplot} Posterior probability
distribution function for the model parameters of equation
(\ref{eq:linear}). {\em Empty contours:} Inference with SL2S galaxies
only. {\em Filled contours:} SL2S + SLACS + LSD lenses. The different
levels represent the $68\%$, $95\%$ and $99.7\%$ enclosed probability
regions.}
\end{figure*}

In this Section we aim to quantify how the mean density slope
$\meangamma$ depends on galaxy properties, and on lookback time.  The
population of ETGs is known to be well-described by two parameters, as
revealed by the existence of the Fundamental Plane relation
\citep{D+D87,Dre++87}.  Two parameters are then probably sufficient to
capture the variation of $\gamma'$ across the population of ETGs.  For
our analysis we focus on stellar mass and effective radius (this
includes also dependencies on stellar mass density, which is believed
to be an important parameter driving $\gamma'$, as discussed above).
Our objective is then to measure the trends in
$\gamma'$ across the
three-dimensional space defined by $(z,M_*,\reff)$.  This is done with
a simple but rigorous Bayesian inference method.  We assume that the
values of the slope $\gamma'$ of our lenses are drawn from a Gaussian
distribution with mean given by
\begin{equation}\label{eq:linear}
%\left<\gamma'\right> = \left<\gamma'_0\right> + \frac{\partial \gamma'}{\partial z}(z-0.3) + \frac{\partial \gamma'}{\partial \log{M_*}}(\log{M_*} - 11.5) + \frac{\partial \gamma'}{\partial \log{\Sigma_*}}(\log{\Sigma_*} - 9)
% \mu_{\gamma'} = \gamma'_0 + \alpha(z-0.3) + \beta(\log{M_*} - 11.5) + \xi\log{(\reff/5)}
\meangamma = \gamma'_0 + \alpha(z-0.3) + \beta(\log{M_*} - 11.5) + \xi\log{(\reff/5)}
\end{equation}
and dispersion $\sigma_{\gamma'}$. 
The stellar mass is in solar units
and the effective radius in kpc. We also assume that individual stellar
masses $M_{*,i}$ are drawn from a parent distribution that we
approximate as a Gaussian:
\begin{equation}
 \pr(M_{*,i}) = 
  \frac{1}{\sigma_{M_{*}}\sqrt{2\pi}}
  \exp{\left[- \frac{\left(\log{M_{*,\mathit{i}}}
                      - \mu_{M_*}^{\mathrm{(Samp)}}(z_{\mathit{i}})\right)}
                    {2\sigma_{M_{*}}^{2\mathrm{(Samp)}}} \right]}.
\end{equation}
To account for selection effects, we allow for a different mean stellar
mass and dispersion for lenses of different surveys. We also let the
mean stellar mass be a function of redshift.  This choice reflects
the clear trend of stellar mass with redshift seen in
\Fref{fig:classevol} for both the SLACS and the SL2S samples, which in
turn is determined by SLACS and SL2S both being magnitude-limited
samples.  The parameter describing the mean stellar mass is then
\begin{equation}
\mu_{M_*}^{\mathrm{(SLACS)}} = \zeta^{\mathrm{(SLACS)}}
  (z_i - 0.2) + \log{M_{*,0}}^{\mathrm{(SLACS)}}
\end{equation}
for SLACS galaxies and
\begin{equation}
\mu_{M_*}^{\mathrm{(SL2S)}} = 
   \zeta^{\mathrm{(SL2S)}}(z_i - 0.5) + \log{M_{*,0}}^{\mathrm{(SL2S)}}
\end{equation}
for SL2S and LSD galaxies.
We assume flat priors on all the model parameters and fit for them with
a Markov chain Monte Carlo following \citet{Kelly07}. 
The stellar masses
considered in this model are those measured in Paper III assuming a
Salpeter IMF. The full posterior probability distribution function is
shown in \Fref{fig:gammap_cornerplot} and the median, 16th and
84th percentile of the probability distribution for the individual
parameters, obtained by marginalizing over the remaining parameters, is
given in \Tref{table:gammap}. The fit is done first with SL2S
galaxies only and then repeated by adding SLACS and LSD lenses. For six
lenses of the SLACS sample \citet{Aug++10} warn that their velocity
dispersions might be significantly incorrect, and we conservatively
exclude them from our fit. These are SSDSJ0029$-$0055, SDSSJ0737$+$3216,
SDSSJ0819$+$4534,
SDSSJ0935$-$0003, SDSSJ1213$+$6708 and SDSSJ1614$+$4522.
\begin{deluxetable*}{cccl}
 \tablecaption{ \label{table:gammap} Linear model with scatter.}
  \tablehead{Parameter & SL2S & SL2S + & Notes \\
            & only & SLACS + LSD & }
  \startdata
  $\log{M_{*,0}}^{\mathrm{(SL2S)}}$ & $11.50_{-0.05}^{+0.05}$ & $11.49_{-0.05}^{+0.05}$ & Mean stellar mass at $z=0.5$, SL2S sample \\ 
$\zeta^{\mathrm{(SL2S)}}$ & $0.35_{-0.33}^{+0.34}$ & $0.38_{-0.26}^{+0.26}$ & Linear dependence of mean stellar mass on redshift, SL2S sample \\ 
$\sigma_{M_*}^{\mathrm{(SL2S)}}$ & $0.25_{-0.04}^{+0.05}$ & $0.23_{-0.04}^{+0.04}$ & Scatter in mean stellar mass, SL2S sample \\ 
$\log{M_{*,0}}^{\mathrm{(SLACS)}}$ & $\cdots$ & $11.59_{-0.03}^{+0.03}$ & Mean stellar mass at $z=0.2$, SLACS sample \\ 
$\zeta^{\mathrm{(SLACS)}}$ & $\cdots$ & $2.35_{-0.39}^{+0.39}$ & Linear dependence of mean stellar mass on redshift, SLACS sample \\ 
$\sigma_{M_*}^{\mathrm{(SLACS)}}$ & $\cdots$ & $0.17_{-0.02}^{+0.02}$ & Scatter in mean stellar mass, SLACS sample \\ 
$\alpha$ & $-0.13_{-0.24}^{+0.24}$ & $-0.31_{-0.10}^{+0.09}$ & Linear dependence of $\gamma'$ on redshift. \\ 
$\beta$ & $0.31_{-0.23}^{+0.23}$ & $0.40_{-0.15}^{+0.16}$ & Linear dependence of $\gamma'$ on $\log{M_*}$. \\ 
$\xi$ & $-0.67_{-0.20}^{+0.20}$ & $-0.76_{-0.15}^{+0.15}$ & Linear dependence of $\gamma'$ on $\log{R_{\mathrm{eff}}}$. \\ 
$\gamma_0$ & $2.05_{-0.06}^{+0.06}$ & $2.08_{-0.02}^{+0.02}$ & Mean slope at $z=0.3$, $\log{M_*} = 11.5$, $R_{\mathrm{eff}} = 5$ kpc \\ 
$\sigma_{\gamma'}$ & $0.14_{-0.03}^{+0.04}$ & $0.12_{-0.02}^{+0.02}$ & Scatter in the $\gamma'$ distribution \\ 

  \enddata
\end{deluxetable*}

By using only the 25 SL2S lenses for which $\gamma'$ measurements are
possible, we are able to detect a trend of 
$\meangamma$ with $\reff$ at
the 3-sigma level and a dependence on $M_*$ at the 1-sigma level: at
fixed $z$ and $M_*$, galaxies with a smaller effective radius have a
steeper density profile. Similarly, at fixed $\reff$, galaxies with a
larger stellar mass have a marginally larger $\gamma'$.  If we add 53 lenses from
SLACS and 4 lenses from the LSD survey, 
the trends with $M_*$ and
$\reff$ are confirmed at a higher significance, and we detect a
dependence of $\meangamma$ on redshift at the 3-sigma level.  
Lower
redshift objects appear to have a steeper 
slope than higher redshift
counterparts at fixed mass and size.  Incidentally, the median value
of $\xi$, the parameter describing the linear dependence of 
$\meangamma$
on $\log{\reff}$, is nearly $-2$ times $\beta$, the parameter
describing the dependence on $\log{M_*}$.  This suggests that
$\meangamma$ grows roughly as $\beta\log{(M_*/\reff^2)}$, which is
equivalent to the stellar mass density.  It appears then that the
dependence of $\gamma'$ on the structure of ETGs can be well
summarized with a dependence on stellar mass density, leaving little
dependence on $M_*$ or $\reff$ individually. This confirms and extends
the trend with surface mass density seen by \citet{Aug++10} and \citet{D+T13}.

We then repeated the fit allowing only for a dependence of $\meangamma$ on
redshift and stellar mass density:
\begin{equation}
\meangamma = \gamma_0 + \alpha(z - 0.3) + \eta(\log{\Sigma_*} - 9.0). 
\end{equation}
This model has one less free parameter with respect to \Eref{eq:linear}.
Our inference on the parameter describing the dependence on $\Sigma_*$ is $\eta = 0.38\pm0.07$, and the scatter in $\gamma'$ is $\sigma_{\gamma'} = 0.12\pm0.02$, the same value measured for the more general model of \Eref{eq:linear}.
This is again suggesting that the dependence of $\gamma'$ on the stellar mass density might be of a more fundamental nature than dependences on mass and size separately.

%-------------------------------------------------------------------------------

\section{Discussion}\label{sect:discuss}

The main result of the previous section is that the ensemble
average total mass density
slope of galaxies of a fixed stellar mass increases with cosmic time
(i.e. decreases with redshift). This trend with redshift is detected
at the $3-\sigma$ confidence level and is in good agreement with
previous results from \citet{Ruf++11} and
\citet{Bol++12}.  

Before discussing the physical interpretation of this result, however,
it is important to emphasize that what we are measuring is how the
population mean density slope changes in the 
$(z,M_*,\reff)$ space within the
population of early-type galaxies, 
and not how $\gamma'$ changes along
the lifetime of an individual galaxy, $\mathrm{d}\gamma'/\mathrm{d}z$. 
In order to infer the latter quantity we need to evaluate the variation of $\gamma'$ along the evolutionary track of the galaxy as this moves in the $(z,M_*,\reff)$ space.
This requires to know how both mass and size of the galaxy change with time, since the slope depends on these parameters.
More formally,
\begin{equation}\label{eq:lagrange}
\begin{split}
\frac{\mathrm{d}\gamma'(z,\log{M_*},\log{\reff})}{\mathrm{d} z} = \\
=\frac{\partial \gamma'}{\partial z} + \frac{\partial \gamma'}{\partial \log M_*}\frac{\mathrm{d}\log M_*}{\mathrm{d}z} + \frac{\partial\gamma'}{\partial \log{\reff}}\frac{\mathrm{d}\log{\reff}}{\mathrm{d}z}.
\end{split}
\end{equation}
In a parallel with fluid mechanics, our description of the population
of galaxies of \Sref{sect:gammap} is Eulerian, while \Eref{eq:lagrange}
is a Lagrangian specification of the change in
time of the mean slope of 
an individual galaxy, 
providing a more
straightforward way to physically understand the evolution of ETGs.

With all these terms entering \Eref{eq:lagrange}, it is no
longer clear if the density slope is indeed getting steeper with time
for individual objects.  In particular, we have observed that
$\gamma'$ depends significantly on stellar mass density (and thus
effective radius). It is then crucial to consider all the terms of the
equation before reaching a conclusion. 
Fortunately this can be done by
combining our measurements 
with results from the
literature.

In the context of our model specified in \Eref{eq:linear},
the partial derivatives introduced above can be identified and
evaluated as follows:
\begin{equation}
\frac{\partial \gamma'}{\partial z} = \alpha =-0.31\pm0.10,
\end{equation}
\begin{equation}
\frac{\partial \gamma'}{\partial \log{M_*}} = \beta = 0.40\pm0.16,
\end{equation}
\begin{equation}
\frac{\partial \gamma'}{\partial \log{\reff}} = \xi = -0.76\pm0.15.
\end{equation}
Note that we are not considering the effects of scatter: we are assuming
that the change in $\gamma'$ is the same as that of a galaxy that evolves 
while staying at the mean $\gamma'$ as it moves through the $(z,M_*,\reff)$
space. 
By doing so, the evolution in the slope that we derive from \Eref{eq:lagrange}
will be representative of the mean change in $\gamma'$ over the population,
while individual objects can have different evolutionary tracks, within the limits allowed by our constraints on $\sigma_{\gamma'}$.

The remaining quantities to be estimated are the rate of mass and size growth.
In the hierarchical
merging picture ETGs are expected to grow in stellar mass with time,
therefore $\mathrm{d} M_*/\mathrm{d} z < 0$.  Observationally, we know
massive early-type galaxies grow at most by a factor of two in stellar
mass since $z=1$ \citep[see, e.g., ][and references
therein]{2013ApJ...771...61L}. Thus we can conservatively take the
mean between zero and 2, even though we will show below that our
conclusion are virtually insensitive to this choice:
\begin{equation}\label{eq:mevol}
\frac{\mathrm{d}\log{M_*}}{\mathrm{d}z} =-0.15\pm0.15.
\end{equation}

The effective radius grows as a result of the growth in mass, but is itself an
evolving quantity at fixed $M_*$ \citep{Dam++11,New++12,Cim++12,Pog++13}:
$\reff = \reff(z,M_*(z))$. {\em We assume that ETGs grow while staying on the
observed $M_*-\reff$ relation at all times.} 
Then we can write
\begin{equation}\label{eq:reffevol}
\frac{\mathrm{d}\log{\reff}}{\mathrm{d}z} = \frac{\partial\log{\reff}}{\partial z} + \frac{\partial\log{\reff}}{\partial\log{M_*}}\frac{\mathrm{d}\log{M_*}}{\mathrm{d}z}
\end{equation}
and use the values measured by \citet{New++12},
$\partial\log{\reff}/\partial z = -0.26\pm0.02$ and
$\partial\log{\reff}/\partial{\log{M_*}} = 0.59\pm0.07$.

\begin{figure}
\includegraphics[width=\columnwidth]{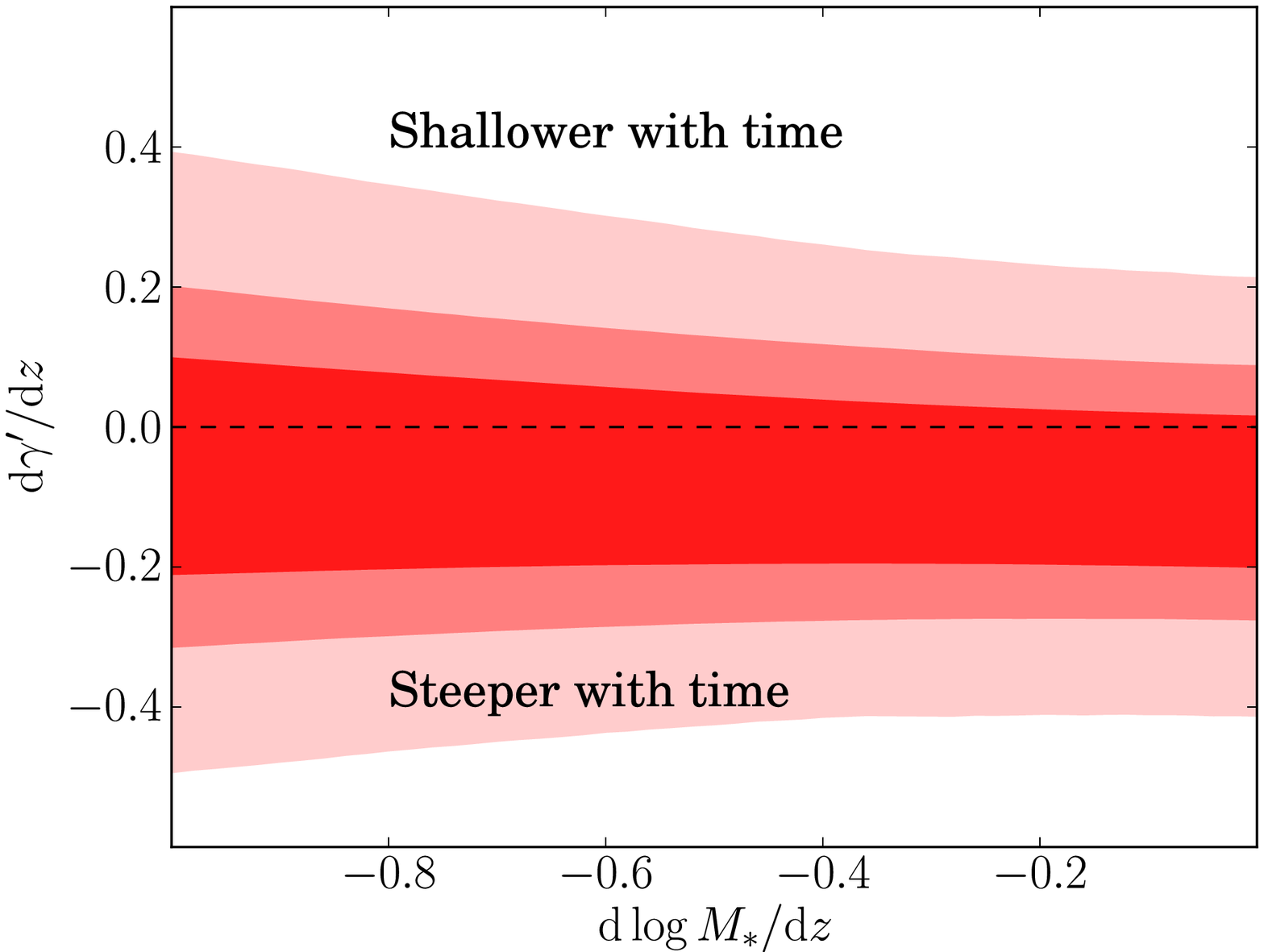}
\caption{\label{fig:total_derivative} Mean intrinsic change of the density slope with redshift of a massive ETG, as a function of its mass growth rate. 
}
\end{figure}

Plugging these values into \Eref{eq:lagrange} we find that
\begin{equation}
\begin{split}
\frac{\mathrm{d}\gamma'}{\mathrm{d}z} = (-0.31\pm0.10) + (0.40\pm0.15)(-0.15\pm0.15) \\
+ (-0.76\pm0.15)[(-0.26\pm0.02) \\ 
+ (-0.15\pm0.15)(0.59\pm0.07)] =-0.10\pm0.12
\end{split}
\end{equation}
Note that $\mathrm{d}\gamma'/\mathrm{d}z$ has little dependence on the mass
growth rate $\mathrm{d}\log{M_*}/\mathrm{d}z$, which is the most poorly known
quantity in this model. To be more quantitative we plot in
\Fref{fig:total_derivative} the total derivative
$\mathrm{d}\gamma'/\mathrm{d}z$ as a function of
$\mathrm{d}\log{M_*}/\mathrm{d}z$, and show that for any plausible value,
spanning over an order of magnitude, the answer is unchanged.  Different
assumptions on the evolution of the size-mass relation do not change
significantly our result.  For instance, \citet{Dam++11} find a more rapid
evolution of $\reff$ than \citet{New++12},  leading to
$\mathrm{d}\gamma'/\mathrm{d}z = 0.06\pm0.15$,
consistent with no change of the total mass
density profile with time.

%%%%%%%%%%%%%%%%%%%%%%%%%%%
%%%%%%%%%%%%%%%%%%%%%%%%%%%

Thus, the key result is that, when considering all the terms of
\Eref{eq:lagrange}, we find that, on average, individual ETGs grow at approximately
constant density slope. 
The observed redshift dependence of $\gamma'$ {\em at fixed mass and size} can then be understood as the result
of the evolution of the size-mass relation and by the dependency of
$\gamma'$ on the stellar mass density. Qualitatively, in this picture
an individual galaxy grows in stellar mass and size so as to decrease
its central stellar mass density. During this process, the slope of
its total mass density profile does not vary significantly.  However
the other galaxies that now find themselves to have the original
stellar mass and effective radius of this galaxy had originally a steeper mass density profile, thus
giving rise to the observed trend in $\partial\gamma'/\partial z$. 

\begin{figure*}
\includegraphics[width=0.9\textwidth]{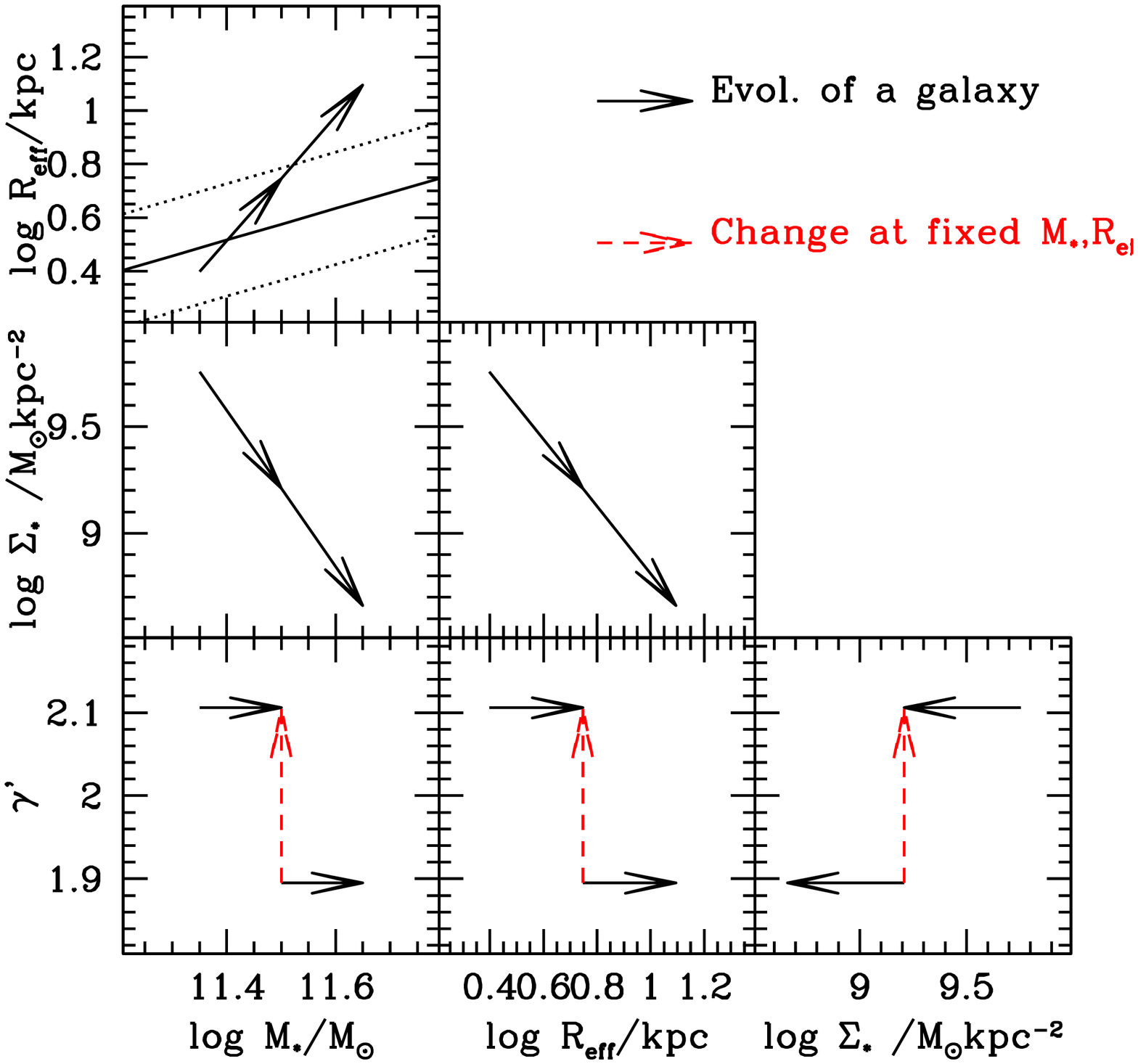}
\caption{\label{fig:toy} Illustration of a scenario consistent with the
observed evolution. The evolutionary tracks of two representative
galaxies between $z=1$ and $z=0$ are shown as solid black arrows, in the
multidimensional parameter space of stellar mass, effective radius,
effective density, and slope of the mass density profile $\gamma'$.
Measured correlations with stellar mass are used to assign the other
parameters as described in the text. 
The solid and dotted lines in the top left panel show the mass-size relation at $z=1$ from \citet{New++12} and the scatter around it.
Even if $\gamma'$ is assumed not to
change for an individual galaxy, $\gamma'$ at fixed stellar mass and size is
observed to increase reflecting the difference in their initial ($z=1$) stellar density, as shown by the red dashed arrows.} 
\end{figure*}

This is illustrated in \Fref{fig:toy}, where we show a possible
scenario consistent with the observations. The evolutionary tracks of
two representative galaxies between $z=1$ and $z=0$ are shown as solid
black arrows, in the multi-dimensional parameter space of stellar
mass, effective radius, effective density, and slope of the mass
density profile $\gamma'$. The two galaxies are chosen so that one has at $z=1$ the same mass and effective radius that the other has at $z=0$.
Mass and size are evolved following \Eref{eq:mevol} and \Eref{eq:reffevol}.
We then assign $\gamma'$ at $z=0$ based on the
observed correlation with size and stellar mass (effectively with
effective stellar mass density, since $\beta\approx-2\xi$) {\em and assume
it does not evolve for an individual galaxy}. 
The apparent evolution of $\gamma'$ at fixed $M_*$ and $\reff$ is consistent with the measured value $\partial \gamma'/\partial z = -0.31\pm0.10$, and is dictated by a difference in the initial stellar density of their progenitors, being larger for the more massive object.

In the context of simple one-parameter stellar profiles (e.g. de Vaucouleurs), this difference in $\gamma'$ at fixed mass and size for galaxies at different redshift must be ascribed to corresponding differences in the underlying dark matter distribution.
The implications of our results for the dark matter profiles of ETGs will be explored in an upcoming paper (Sonnenfeld et al., in prep.).

%%%%%%%%%%%%%%%%%%%%%%%%%%%%%
%%%%%%%%%%%%%%%%%%%%%%%%%%%%%

An important assumption at the basis of our analysis is that scaling relations of $\gamma'$ with mass and size measured at low redshift can be used to predict the evolution of the slope for higher redshift objects.
This assumption holds well if the evolutionary tracks of higher redshift galaxies stay on parts of the parameter space probed by the lower redshift systems. 
To first approximation this seems to be the case for the galaxies in our sample.
\Fref{fig:mreffz0} shows the positions of our lenses in the $M_*-\reff$ space, where the effective radius of each object is renormalized by the average $\reff$ of galaxies at its redshift.
Under our assumptions, objects evolve along lines parallel to the mass-size relation (dashed line) towards higher masses.
There is significant overlap between the high-$z$ SL2S-LSD sample and the lower redshift SLACS sample, implying that SLACS galaxies are informative on the evolution in $\gamma'$ of SL2S-LSD objects.
Differently, one could rely on extrapolations of the scaling relations for $\gamma'$.

A more quantitative explanation of our findings would require a detailed
comparison with theoretical model and is beyond the scope of this
work. However, we can check at least qualitatively how our result
compares with published predictions. \citet{NTB09} studied the impact
of dissipationless (dry) mergers on $\gamma'$ finding that for an
individual galaxy the slope tends to get shallower with
time. \citet{JNO++12} looked at the evolution in the slope on nine
ETGs in cosmological simulations, finding no clear trend in the
redshift range explored by our data. Their simulations include both
dry and dissipational (wet) mergers. \citet{Rem++13} examined
simulated ETGs in a cosmological framework and in binary mergers. They
found slopes that become shallower in time, asymptotically approaching
the value $\gamma'\approx2.1$ as observed in our data. They also
detected a correlation between the amount of in-situ star formation
and slope, with $\gamma'$ being larger in systems that experienced
more star formation events.  Finally, \citet{Dub++13} produced zoomed
cosmological simulations of ETGs with or without AGN feedback. They
found that the total density slope becomes steeper with time. They
also observed that galaxies with strong AGN feedback have a shallower
profile than systems with no AGN feedback and interpreted this result
with the AGN shutting off in-situ star formation.  Qualitatively, our
data is not in stark contrast with any of these models.

A more quantitative comparison is required to find out whether the
models work in detail. This is left for future work. The combination
of constraints from the evolution of the size stellar mass relation
obtained via traditional studies of large samples of ETGs, and our own
detailed measurements of the evolution of their internal structure
should provide a stringent test for evolutionary models of ETGs, and
thus help us improve our understanding of the baryonic and dark matter
physics relevant at kpc scales.

%-------------------------------------------------------------------------------

\section{Summary and Conclusions}\label{sect:concl} 

We have presented  spectroscopic observations from the Keck, VLT, and
Gemini Telescopes of a sample of 53 lenses and lens candidates from the
SL2S survey.  We measured stellar velocity dispersions for 47 of them,
and redshifts of both lens and background source for 35 of them.  36
systems are confirmed grade A lenses and 25 of these were able to be
used for a lensing and stellar dynamics analysis.  We have shown how
spectroscopic observations can be used in combination with ground-based
imaging with good seeing ($\sim 0.7''$) to confirm gravitational lens
candidates by the presence of multiply imaged emission lines from the
lensed background source.  We have also shown how SL2S lenses are
comparable with lenses from other surveys in terms of their size, mass
and velocity dispersion, and lie on the same $M_*-\reff$ relation as non-lens galaxies.

By fitting a power-law density profile ($\rho(r) \propto
r^{-\gamma'}$) to the lensing and stellar kinematics data of SL2S,
SLACS and LSD lenses we measured the dependence of $\gamma'$ on
redshift, stellar mass and galaxy size, over the ranges $z \approx
0.1-1.0$, $\log M_*/M_\odot \approx 11 - 12$, R$_{\rm eff}=1-20$kpc.

Our main results can be summarized as follows:

\begin{enumerate}

\item  
In the context of power-law models for the density profile $\rho_{\rm tot}\propto r^{-\gamma'}$,
the (logarithmic) density slope
$\gamma'$ of the SL2S lenses is approximately -- but not exactly
-- that of a single isothermal sphere ($\gamma'=2$), consistent with
previous studies of lenses in different samples.  This can be
understood as the result of the combination of a stellar mass density
profile that falls off more steeply than the dark matter halo. The
relative scaling of the two conspires to produce the power law
index close to isothermal (``bulge-halo'' conspiracy).

\item At a given redshift, the 
mass density slope $\gamma'$ depends on the
surface stellar mass density $\Sigma_*=M_*/2\reff^2$, in the sense
that galaxies with denser stars also have steeper total mass density
profiles ($\partial \gamma' / \partial \log \Sigma_* = 0.38\pm0.07$).

\item At fixed $M_*$ and $\reff$, $\meangamma$ depends on redshift, in the
sense that galaxies at a lower redshifts have on average a steeper
average slope ($\partial \gamma'/ \partial z = -0.31\pm 0.10$).

\item Once the dependencies of $\gamma'$ on redshift and surface stellar
mass density are taken into account, less than 6\% intrinsic scatter is
left ($\sigma_\gamma'=0.12\pm0.02$). 

\item 
The average redshift evolution of $\gamma'$ for an individual galaxy is
consistent with zero: $\mathrm{d}\gamma'/\mathrm{d}z=-0.10\pm0.12$. This result is
obtained by combining our measured dependencies of $\meangamma$ on
redshift stellar mass and effective radius with the observed evolution
of the size stellar mass relation taken from the literature.

\end{enumerate}

The key result of this work is that the dependency of $\meangamma$ on
redshift and stellar mass density does not imply that massive
early-type galaxies change their mass density profile over the
second half of the lifetime. In fact, at least qualitatively, the
observed dependencies can be understood as the results of two
effects. Individual galaxies grow in stellar mass and decrease in
density over the redshift range 1 to 0,
while apparently largely preserving their
total mass density profiles. This could be explained by the addition of
stellar mass in the outer part of the galaxies in quantities that are
sufficient to explain the decrease in stellar mass density but
insufficient to alter the total mass density profile, since the
regions are already dark matter dominated. As shown by
\citet{Nip++12}, the growth in size during this period is slow enough
that it could perhaps be explained by the the infall of dark matter
and stars via a drizzle of minor mergers, with material of decreasing
density, tracking the decreasing cosmic density. This process needs to
happen while substantially preserving the total mass density
profile. 

Alternatively, the evolution at constant slope can be interpreted as the combined effect of the decrease in stellar mass density and a variation in the dark matter profile (either a steepening or a decrease of the central dark matter distribution).
The latter effect would be responsible for the term $\partial\gamma'/\partial z$.

Checking whether these scenarios can work quantitatively
requires detailed comparisons with theoretical calculations, which are
beyond the scope of this paper.

The second important result of this work is that the total mass
density profile of early-type galaxies depends on their stellar mass
density, with very little scatter. 
Qualitatively this makes sense, as
we expect that the more concentrated stellar distributions should have
been able to contract the overall profile the most. Presumably this
difference may trace back to differences in past star formation
efficiency or merger history. Therefore, the tightness of the observed
correlation should provide interesting constraints on these crucial
ingredients of our understanding of early-type galaxies. 

%-------------------------------------------------------------------------------

\acknowledgments

We thank our friends of the SLACS and SL2S collaborations for many
useful and insightful discussions over the course of the past years.
We thank V.N.~Bennert and M.~Bradac for their help in our observational campaign.
TT thanks S.W.~Allen and B.~Poggianti for
useful discussions.
RG acknowledges support from the Centre National des Etudes Spatiales
(CNES).
PJM acknowledges support from the Royal
Society in the form of a research fellowship. 
TT acknowledges support from the NSF through CAREER award NSF-0642621, and from
the Packard Foundation through a Packard Research Fellowship.
This research is based on XSHOOTER observations made with ESO Telescopes at the Paranal
Observatory under programme IDs 086.B-0407(A) and 089.B-0057(A).
This research is based on observations obtained with MegaPrime/MegaCam, a joint project of CFHT
and CEA/DAPNIA, and with WIRCam, a joint project of CFHT, Taiwan,
Korea, Canada and France, at the Canada-France-Hawaii Telescope (CFHT) which is operated
by the National Research Council (NRC) of Canada, the Institut National des
Sciences de l'Univers of the Centre National de la Recherche Scientifique
(CNRS) of France, and the University of Hawaii. This work is based in part on
data products produced at TERAPIX and the Canadian Astronomy Data Centre.
The authors would like to thank S. Arnouts, L. Van waerbeke and G. Morrison for giving access to the WIRCam data collected in W1 and W4 as part of additional CFHT programs. We are particularly thankful to Terapix for the data reduction of this dataset.
This research is supported by NASA through Hubble Space Telescope programs
GO-10876, GO-11289, GO-11588 and in part by the National Science Foundation
under Grant No. PHY99-07949, and is based on observations made with the
NASA/ESA Hubble Space Telescope and obtained at the Space Telescope Science
Institute, which is operated by the Association of Universities for Research in
Astronomy, Inc., under NASA contract NAS 5-26555, and at the W.M. Keck
Observatory, which is operated as a scientific partnership among the California
Institute of Technology, the University of California and the National
Aeronautics and Space Administration. The Observatory was made possible by the
generous financial support of the W.M. Keck Foundation. The authors wish to
recognize and acknowledge the very significant cultural role and reverence that
the summit of Mauna Kea has always had within the indigenous Hawaiian
community.  We are most fortunate to have the opportunity to conduct
observations from this mountain.

%-------------------------------------------------------------------------------

\bibliographystyle{apj}
\bibliography{references}

%-------------------------------------------------------------------------------

\end{document}